\pdfoutput=1

\documentclass[11pt,twoside,a4paper,cmspaper,final,collab]{cms-tdr}
\def\svnVersion{177013}\def\cmsCernNo{2012-273}\def\cmsCernDate{\today}\def\cmsMessage{Submitted to the Journal of High Energy Physics}

\begin{document}\cmsNoteHeader{SUS-12-006}

\hyphenation{had-ron-i-za-tion}
\hyphenation{cal-or-i-me-ter}
\hyphenation{de-vices}

\RCS$Revision: 149587 $
\RCS$HeadURL: svn+ssh://svn.cern.ch/reps/tdr2/papers/SUS-12-006/trunk/SUS-12-006.tex $
\RCS$Id: SUS-12-006.tex 149587 2012-09-27 17:29:08Z alverson $

\newcommand{\fullLumi}{4.98\fbinv}  % fbinv has xspace
\newcommand{\fullLumiError}{$4.98 \pm 0.11\fbinv$\xspace}  %error copied from SUS-11-010 paper

\newcommand{\slep}{\ensuremath{\widetilde{\ell}}\xspace}
\newcommand{\xslep}{\ensuremath{x_{\slep}}\xspace}
\newcommand{\snu}{\ensuremath{\widetilde{\nu}}\xspace}
\newcommand{\mchi}{\ensuremath{{m_{\chiz_2}=m_{\chipm_1}}}\xspace}

\newcommand{\Irel}{\ensuremath{I_\text{rel}}\xspace} %11-013 variable
\newcommand{\W}{\PW\xspace} % \Z is defined by default but not \W; there is \PW in PENNAMES
\newcommand{\ZZ}{\cPZ\cPZ\xspace}
\newcommand{\bjet}{\ensuremath{\cPqb\text{-jet}}\xspace}
\newcommand{\bjetnohyphen}{\ensuremath{\cPqb\ \text{jet}}\xspace}
\newcommand{\bjets}{\ensuremath{\cPqb\text{-jets}}\xspace}
\newcommand{\bjetsnohyphen}{\ensuremath{\cPqb\ \text{jets}}\xspace}

\newcommand{\bquark}{\ensuremath{\cPqb\text{ quark}}\xspace}
\newcommand{\cquark}{\ensuremath{\cPqc\text{ quark}}\xspace}
\newcommand{\udsgparton}{\ensuremath{\cPqu\cPqd\cPqs\cPg\text{ parton}}\xspace}

\newcommand{\CTEQ} {{\textsc{cteq}}\xspace}

\newcommand{\vecMET}{\ensuremath{\vec{E}_{\mathrm{T}}^{\text{miss}}}\xspace}

\newcommand{\MT}{\ensuremath{M_\mathrm{T}}\xspace}
\newcommand{\mdil}{\ensuremath{M_{\ell\ell}}\xspace}
\newcommand{\mjj}{\ensuremath{M_{jj}}\xspace}

\newcommand{\ST}{\ensuremath{S_{\mathrm{T}}}\xspace}

\newcommand{\zjets}{\ensuremath{\Z+\text{jets}}\xspace}
\newcommand{\wwjets}{\ensuremath{\W\W+\text{jets}}\xspace}
\newcommand{\wzjets}{\ensuremath{\W\Z+\text{jets}}\xspace}
\newcommand{\gjets}{\ensuremath{\gamma+\text{jets}}\xspace}

\newcommand{\zzmet}{$\Z\Z+\MET$\xspace}
\newcommand{\wzmet}{$\W\Z+\MET$\xspace}
\newcommand{\wzzmet}{$\W\Z/\Z\Z+\MET$\xspace}
\newcommand{\cls}{CL$_\text{s}$\xspace}

\newcommand{\fix}[1]{{\bf <<< #1 !!! }}% Bob's favorite fix macro

\cmsNoteHeader{SUS-12-006} % This is over-written in the CMS environment: useful as preprint no. for export versions
\title{Search for electroweak production of charginos and neutralinos using leptonic
final states in pp collisions at $\sqrt{s} = 7\TeV$}

\date{\today}

\abstract{
The 2011 dataset of the CMS experiment, consisting of an integrated
luminosity of 4.98\fbinv of pp collisions at
$\sqrt{s} = 7$\TeV, enables expanded searches for direct electroweak
pair production of charginos and neutralinos in supersymmetric models
as well as their analogs in other models of new physics.  Searches
sensitive to such processes, with decays to final states that contain
two or more leptons, are presented.
Final states with three leptons, with a same-sign lepton
pair, and with an opposite-sign lepton pair in conjunction with two
jets, are examined.  No excesses above the standard model expectations
are observed. The results are used in conjunction with previous
results on four-lepton final states to exclude a range of chargino and
neutralino masses from approximately 200 to 500\GeV in the context of models
that assume large branching fractions of charginos and neutralinos to leptons
and vector bosons.}

\hypersetup{%
pdfauthor={CMS Collaboration},%
pdftitle={Search for electroweak production of charginos and neutralinos using leptonic final states in pp collisions at sqrt(s) = 7 TeV},%
pdfsubject={CMS},%
pdfkeywords={CMS, physics, supersymmetry}}

\maketitle %maketitle comes after all the front information has been supplied

\section{Introduction}
\label{introduction}

Many searches for physics beyond the standard model (BSM) performed by
experiments at the CERN Large Hadron Collider (LHC) have focused on
models with cross sections dominated by the production of new heavy
strongly interacting particles, with final states characterized by
large hadronic activity.  These searches are well justified since
strongly interacting particles can be produced with large cross
sections and hence be observable with early LHC data. In the context
of supersymmetry (SUSY)
\cite{Golfand:1971iw,Ramond:1971gb,Neveu:1971rx,Neveu:1971iv,Volkov:1973ix,Wess:1973kz,Wess:1974tw},
such models lead mainly to the production of the strongly interacting
squarks and gluinos, the SUSY partners of the quarks and gluons. In
contrast, in this paper we describe searches for BSM physics dominated
by the direct electroweak production of particles that might not yield
large hadronic activity, and that may therefore have eluded detection
in early searches.  This signature characterizes SUSY models with
pair-production of electroweak charginos $\chipm$ and neutralinos
$\chiz$, mixtures of the SUSY partners of the gauge bosons and Higgs
bosons.  Depending on the mass spectrum, the charginos and neutralinos
can have significant decay branching fractions (BF) to leptons or vector
bosons, resulting in final states that contain either on-shell vector
bosons or three-lepton states with continuous pair-mass
distributions~\cite{Dicus:1983cb,Chamseddine:1983eg,Chamseddine:1983eg,Baer:1985at,Nath:1987sw,Baer:1994nr,Baer:1995va}.
In either case, neutrino(s) and two stable lightest-SUSY-particle
(LSP) dark-matter candidates are produced, which escape without
detection and lead to large missing transverse energy \MET in the
event.

\begin{figure}[!h]
\begin{center}
\subfigure[]{\includegraphics[width=0.4\textwidth]{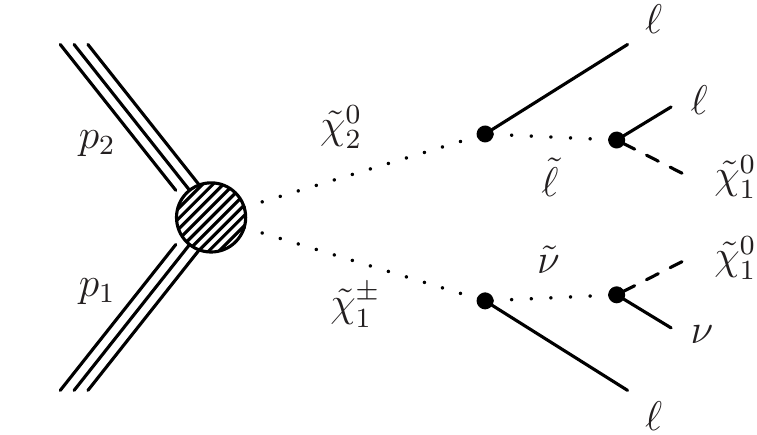} }
\subfigure[]{\includegraphics[width=0.4\textwidth]{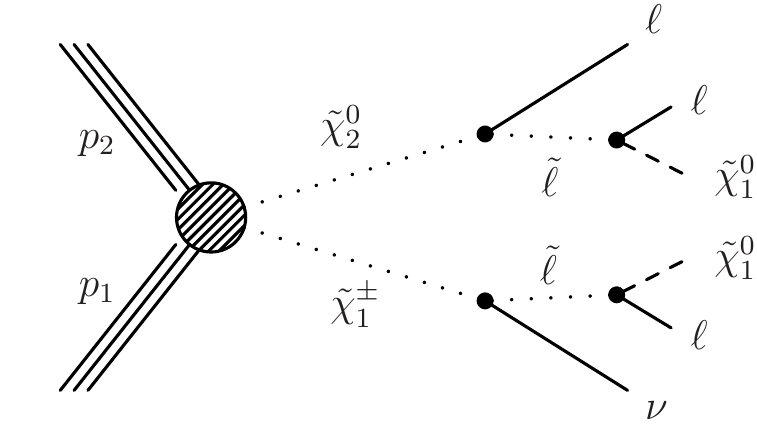} }
\caption{Diagrams of chargino-neutralino pair production in proton-proton
collisions followed by decays leading to a final state with three
leptons, two LSPs, and a neutrino. For left-handed sleptons (with
accompanying sneutrinos), both diagrams exist, and for each diagram
there is an additional diagram with
$\chiz_2\rightarrow\ell\,\slep\rightarrow\ell\,\ell\,\chiz_1$ replaced
by $\chiz_2\rightarrow\snu\,\nu\rightarrow\nu\,\nu\,\chiz_1$.  Thus
only 50\% of produced pairs results in three leptons.  For
right-handed sleptons, only the right diagram exists, and 100\% of
produced pairs result in three leptons. In these diagrams and those of
Fig.~\ref{fig:charginos-wz}, dotted lines represent unstable
intermediate states, and the dashed lines represent the LSP.
\label{fig:charginos-slep}
}
\end{center}
\end{figure}

\begin{figure}[!h]
\begin{center}
\subfigure[]{\includegraphics[width=0.4\textwidth]{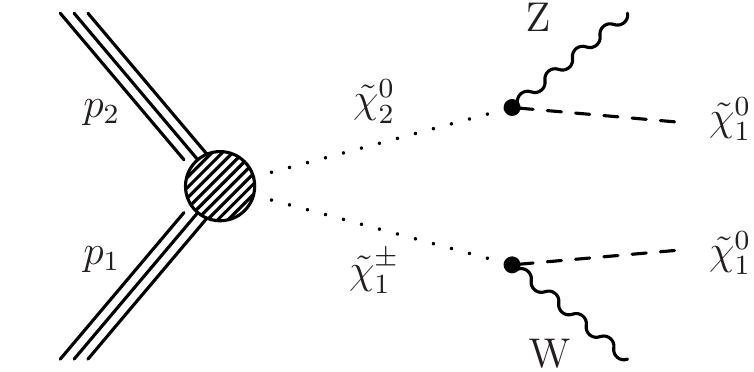} }
\subfigure[]{\includegraphics[width=0.4\textwidth]{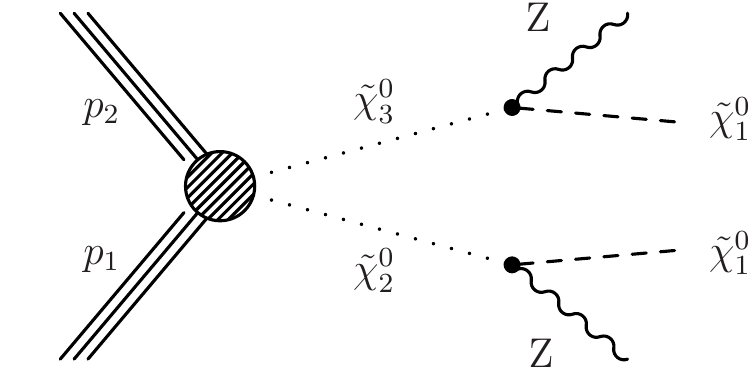} }
\caption{Diagrams of chargino-neutralino and neutralino-neutralino
pair production in proton-proton collisions followed by decay to
on-shell \W or \Z bosons and LSPs.
\label{fig:charginos-wz}
}
\end{center}
\end{figure}

In this paper, we present several dedicated searches for
chargino-neutralino pair production.  The data, corresponding to an
integrated luminosity of \fullLumiError \cite{CMS-PAS-SMP-12-008} of
proton-proton collisions at $\sqrt{s} =7\TeV$, were collected by the
Compact Muon Solenoid (CMS) experiment at the LHC in 2011. Even with
the smaller cross sections of electroweak production, this data sample
is sufficient to probe the production of charginos and neutralinos
with masses well beyond existing
constraints~\cite{aleph,Heister:2002jca,delphi,l3,opal,Aaltonen:2008pv,Abazov:2009zi,atlas-trilepton}.
Since LHC studies have as yet found no evidence for new strongly
interacting particles, we focus on scenarios in which such particles
do not participate, and in which the final states are rich in leptons
produced via intermediate states including sleptons (SUSY partners of
the leptons, including sneutrinos, partners of neutrinos).  These
scenarios include cases such as those shown in
Figs.~\ref{fig:charginos-slep} and \ref{fig:charginos-wz}, which are
labeled using SUSY nomenclature, though the interpretation naturally
extends to other BSM models. In the SUSY  nomenclature,
$\chiz_1$ is the lightest neutralino,
presumed to be the LSP, and $\chiz_2$ and $\chiz_3$
are heavier neutralinos; $\chipm_1$ is the lightest chargino.  In
Fig.~\ref{fig:charginos-slep} the slepton mass $m_{\slep}$ is less
than the masses $m_{\chiz_2}$ and $m_{\chipm_1}$, while in
Fig.~\ref{fig:charginos-wz} it is greater, and the mass difference
between the LSP and the next-lightest chargino or neutralino is large
enough to lead to on-shell vector bosons.  In addition to the
dedicated searches, we leverage the results of some previous CMS SUSY
searches~\cite{Chatrchyan:2011ff,SUS-11-013-paper,SUS-11-021-paper,SUS-11-010},
either by interpreting the previous results directly in the context of
the scenarios in Figs.~\ref{fig:charginos-slep} and
\ref{fig:charginos-wz}, or by modifying the previous studies so that
they target electroweak, rather than strong, production processes.
Throughout this paper, ``lepton'' refers to a charged lepton; in
specified contexts, it refers more specifically to an experimentally
identified electron or muon.

To quantify our results, we present them in the context of simplified
model spectra (SMS)
\cite{Knuteson:2006ha,ArkaniHamed:2007fw,Dube:2008kf,Alwall:2008ag,Alwall:2008va,Alves:2011wf,Alves:2011sq,Papucci:20113g}.
SUSY models with bino-like $\chiz_1$ and wino-like $\chiz_2$ and
$\chipm_1$ lead to the SMS trilepton signature of
Fig.~\ref{fig:charginos-slep}, and motivate the simplifying assumption
that the latter two gauginos have similar masses as a result of
belonging to the same gauge group multiplet.  We thus set
$m_{\chiz_2}=m_{\chipm_1}$, and present results as a function of this
common mass and the LSP mass $m_{\chiz_1}$.  The results for
Fig.~\ref{fig:charginos-slep} depend also on the mass
$m_{\slep}$ of the intermediate slepton (if left-handed, taken to be the same for
its sneutrino $\snu$), parametrized in terms of a variable $\xslep$ as
\begin{equation}
  m_{\slep} = m_{\chiz_1} +  \xslep\, ( m_{\chipm_1} - m_{\chiz_1}\,),
\end{equation}
where $0<\xslep<1$. We present results for $\xslep$ equal to 0.5
(i.e., the slepton mass equal to the mean of the LSP and chargino
masses). In some cases we also present results for $\xslep=$0.25 and
0.75.

The interpretation of the result may further depend on whether the
sleptons are the SUSY partner $\slep_L$ or $\slep_R$ of left-handed or
right-handed leptons.  We consider two limiting cases.  In one case,
$\slep_R$ does not participate while $\slep_L$ and $\snu$ do: then
both diagrams of Fig.~\ref{fig:charginos-slep} exist, and the chargino
and neutralino decay to all three lepton flavors with equal
probability.  Furthermore, two additional diagrams with
$\chiz_2\rightarrow\ell\,\slep\rightarrow\ell\,\ell\,\chiz_1$ replaced
by $\chiz_2\rightarrow\snu\,\nu\rightarrow\nu\,\nu\,\chiz_1$ reduce
the fraction of three-lepton final states by 50\%.  In the second
case, in which $\slep_R$ participates while $\slep_L$ and $\snu$ do
not, only the diagram of Fig.~\ref{fig:charginos-slep}(b) exists, and
there is no 50\% loss of three-lepton final states.  Because the
$\slep_R$ couples to the chargino via its higgsino component, chargino
decays to $\slep_R$ strongly favor the $\tau$ as the lepton.  For the
leptonic decay products, we thus consider primarily two flavor
scenarios:
\begin{itemize}
\item The ``flavor-democratic'' scenario: the chargino ($\chipm_1$) and
neutralino ($\chiz_2$) both decay with equal probability into all
three lepton flavors, as expected for $\slep_L$;
\item The ``$\tau$-enriched'' scenario: the chargino decays exclusively to
$\tau$ leptons as expected for $\slep_R$, while the neutralino decays
democratically.
\end{itemize}
With the selection criteria used in this paper, we have only limited
sensitivity to a third scenario: the ``$\tau$-dominated'' scenario in
which the chargino and neutralino both decay only to a $\tau$ lepton.

We place limits on the pair production cross section times branching
fraction in the above scenarios.  In additional interpretations given
below in terms of bounds on masses within SMS, the 50\% branching
fraction to three leptons is taken into account when appropriate in
$\slep_L$ cases.  For $\xslep=0.5$, the kinematic conditions for the
processes of Fig.~\ref{fig:charginos-slep} are identical for $\slep_L$
and $\slep_R$, and the respective limits are trivially related.  For
other values of $\xslep$ (0.25 and 0.75), differences in experimental
acceptance may alter the relationship.

For results based on the diagrams of Fig.~\ref{fig:charginos-wz}, we
assume that sleptons are too massive to participate, so that the
branching fractions to vector bosons are 100\%.  Even with such an
assumption, there is little sensitivity to the \ZZ channel of
Fig.~\ref{fig:charginos-wz}(b) in the context of models such as the
minimal supersymmetric extension of the standard model (MSSM), where
neutralino pair production is suppressed relative to
neutralino-chargino production.  Rather, for the \ZZ signature, we
consider a specific gauge-mediated supersymmetry breaking (GMSB)
\Z-enriched higgsino
model~\cite{Matchev:1999ft,Meade:2009qv,ref:ewkino} that enhances the
$\ZZ + \MET$ final state.

Following a description of the data collection and reconstruction
procedures in Section~\ref{detector}, Section~\ref{trilepton}
describes searches specifically aimed at the three-lepton final state
of Fig.~\ref{fig:charginos-slep}.  Kinematic observables that can
distinguish signal from background
include~\cite{Matchev:1999nb,Baer:1999bq,Matchev:1999yn,Barger:1998hp}
$\MET$, the invariant mass $\mdil$ of the opposite-sign leptons, and
the transverse mass \MT formed from one lepton and the \MET.  A
three-lepton search using $\MET$ is presented in
Section~\ref{trilepton-broad}, while a complementary approach using
$\mdil$ and $\MT$ is presented in Section~\ref{trilepton-targeted}.
In these three-lepton searches, the leptons selected are electrons and
muons. Sensitivity to $\tau$ leptons arises only through their
leptonic decays.

The three-lepton searches lose sensitivity when the probability to
detect the third lepton becomes low.  In Section~\ref{dilepton}, we
describe a search based on exactly two reconstructed leptons with the
same electric charge (same sign), which extends the sensitivity to the
processes of Fig.~\ref{fig:charginos-slep}.  This study, a
modification of the CMS search for SUSY described in
Ref.~\citen{SUS-11-010}, includes hadronically decaying $\tau$ leptons
in addition to electrons and muons.  Section~\ref{diboson} describes a
search for the on-shell \W and \Z boson production processes of
Fig.~\ref{fig:charginos-wz}. This study is a modification of the CMS
search for SUSY in the
\Z boson plus jets and \MET channel~\citen{SUS-11-021-paper}.

Section~\ref{conclusion} presents an interpretation of these searches,
in some cases combining several together, and including the
four-lepton results of Ref.~\citen{SUS-11-013-paper}.  Results of
related searches have also been recently reported by the ATLAS
collaboration~\cite{atlas1208.3144,atlas1208.2884}.

Finally, Appendix~\ref{efficiency} provides a parametrized function
for the detection efficiency of physics objects used in the analysis
in Section~\ref{trilepton-targeted}.  This function will enable
estimation of sensitivities for BSM models not considered in this paper
that yield three leptons in the final state.

\section{Detector, online selection, and object selection}
\label{detector}

The online event selections (trigger) and further offline object
selections closely follow those described in
Ref.~\citen{SUS-11-013-paper}, and are briefly summarized here.
Exceptions are noted below in the sections specific to each analysis.

The central feature of the CMS apparatus is a superconducting
solenoid, of 6\unit{m} internal diameter, providing a magnetic field of
3.8\unit{T}. Within the field volume are a silicon pixel and strip
tracker, a crystal electromagnetic calorimeter (ECAL), and a
brass/scintillator hadron calorimeter. Muons are measured in
gas-ionization detectors embedded in the steel return yoke. Extensive
forward calorimetry complements the coverage provided by the barrel
and endcap detectors. A more detailed description can be found in
Ref.~\citen{:2008zzk}.

CMS uses a right-handed coordinate system, with the origin at the
nominal interaction point, the $x$ axis pointing to the center of the
LHC, the $y$ axis pointing upwards (perpendicular to the plane of the
LHC ring), and the $z$ axis along the counterclockwise-beam
direction. The polar angle $\theta$ is measured from the positive $z$
axis, and the azimuthal angle $\phi$ (in radians) is measured in the
$x$-$y$ plane. The pseudorapidity $\eta$ is a transformation of the
polar angle defined by $\eta = -\ln [\tan (\theta/2)]$.

Events from pp interactions must satisfy the requirements of a
two-level trigger system. The first level performs a fast selection
for physics objects (jets, muons, electrons, and photons) above
certain thresholds.  The second level performs a full event
reconstruction. Events in this analysis are primarily selected using
double-lepton triggers that require at least one electron or muon with
transverse momentum $\pt > 17\GeV$, and another with $\pt > 8\GeV$,
with $|\eta|<2.5$ for electrons and $|\eta|<2.4$ for muons.  For
channels involving $\tau$ leptons, triggers are used that rely on
significant hadronic activity and $\MET$, in addition to the presence
of a single lepton or two hadronic $\tau$
candidates~\cite{SUS-11-010}. Additional triggers are used for
calibration and efficiency studies.

Simulated event samples are used to study the characteristics of
signal and standard model (SM) background.  Most of the simulated
event samples are produced with the \MADGRAPH
5.1.1~\cite{Maltoni:2002qb,Alwall:2011uj} event generator, with parton
showering and hadronization performed with the \PYTHIA
8.1~\cite{Sjostrand:2007gs} program.  Signal samples are generated
with
\PYTHIA 6.424~\cite{Sjostrand:2007gs}.
The samples are generated using the \CTEQ 6L1~\cite{PhysRevD.78.013004}
parton distribution functions. For the diboson backgrounds,
\MCFM \cite{Campbell:2011bn} samples are used to help assess the theoretical
uncertainties on the simulated samples.  For the simulated SM samples,
we use the most accurate calculations of the cross sections available,
generally with next-to-leading order (NLO)
accuracy~\cite{Beenakker:1999xh,Beenakker:1999xhErr,bib-NLO-NLL}.
The files specifying the SUSY signal model parameters are generated
according to the SUSY Les Houches accord~\cite{Skands:2003cj}
standards with the \ISAJET program~\cite{Baer:1993ae}, with cross
sections calculated in \PYTHIA to leading order and NLO corrections
calculated using \PROSPINO 2.1 \cite{Beenakker:1996ed}.  Depending on
the simulated sample, the detector response and reconstruction are
modeled either with the CMS fast simulation
framework~\cite{Abdullin:1328345}, or with the \GEANTfour~\cite{Geant}
program, followed by the same event reconstruction as that used for data.

Events are reconstructed offline using the particle-flow (PF)
algorithm~\cite{PFT-10-004,PFT-08-001}, which provides a
self-consistent global assignment of momenta and energies.  Details of
the reconstruction and identification are given in
Refs.~\citen{EGM-10-004,MUO-10-004} for electrons and muons.
Leptonically decaying $\tau$ leptons are included in the selection of
electrons or muons.  In the same-sign dilepton search, hadronic $\tau$
lepton decays are identified with the ``hadrons-plus-strips''
algorithm~\cite{SUS-11-010,PFT-10-XXX}.  This algorithm combines PF
photons and electrons into strips (caused by azimuthal bending of an
electromagnetic shower in the CMS magnetic field) in order to
reconstruct neutral pions. The neutral pions are combined with charged
hadrons to reconstruct exclusive hadronic $\tau$ decay topologies.  In
the four-lepton results from Ref.~\citen{SUS-11-013-paper} used in the
interpretations in Section~\ref{conclusion}, hadronic $\tau$
candidates are identified as isolated tracks with associated ECAL
energy deposits consistent with those from neutral pions.

We consider events that contain electrons, muons, and (for a subset of
the searches, as specified above) hadronically decaying $\tau$
leptons, each associated with the same primary vertex.  Offline
requirements on the lepton $\PT$ and $\eta$ are described in the
analysis-specific sections below.  To reduce contamination due to
leptons from heavy-flavor decays or misidentified hadrons in jets, an
isolation criterion is formed by summing the track $\pt$ and
calorimeter $\ET$ values in a cone of $\Delta R = 0.3$ (0.4 for
electrons in the three-lepton$+\MET$ search) around the lepton, where
$\Delta R = \sqrt{(\Delta\phi)^2 + (\Delta\eta)^2}$.  The candidate
lepton is excluded from the isolation sum.  This sum is
divided by the lepton's $\pt$ to obtain the isolation ratio $\Irel$,
which is required to be less than 0.15.

Jets are reconstructed with the anti-$k_{\mathrm{T}}$ clustering
algorithm~\cite{Cacciari:2008gp} with a distance parameter of 0.5. The
jet reconstruction is based on PF objects. With exceptions noted
below, jets are required to have $|\eta| < 2.5$ and $\pt > 40\GeV$ and
to be separated from any lepton satisfying the analysis selection by
$\Delta R > 0.3$.  Where applicable to suppress background from heavy
flavors, we identify jets with $\cPqb$ quarks (referred to throughout
as ``\bjetsnohyphen'') by using the CMS ``track-counting
high-efficiency algorithm'' (TCHE)~\cite{btag-11-001}, which provides
a \bjet tagging efficiency of 76\% (63\%) with a misidentification
rate of 13\% (2\%) for the loose (medium) working point.

Events with an opposite-sign same-flavor (OSSF) dilepton (i.e.,
dielectron or dimuon) with invariant mass below 12\GeV are rejected,
to exclude quarkonia resonances, low-mass continuum, and photon
conversions.

\section{Searches in the three-lepton final state}
\label{trilepton}

For the searches in the three-lepton final state, we use reconstructed
leptons identified as electrons and muons; any sensitivity to $\tau$
leptons comes indirectly through their leptonic decays.  The main SM
backgrounds in the three-lepton final state are from $\W\Z$ production
with three genuine isolated leptons that are ``prompt'' (created at
the primary vertex), and from $\ttbar$ production with two such
leptons and a third particle identified as such but that is
``non-prompt'' (created at a secondary vertex, as from a heavy-flavor
decay) or not a lepton.  We consider two complementary variants of
this search.  The first uses the missing transverse energy $\MET$
directly, and has slightly better sensitivity than the second when the
difference between $\mchi$ and the LSP mass $m_{\chiz_1}$ is large.
The second search uses $\MET$ indirectly through the transverse mass
$\MT$, which is particularly effective in discriminating background
from leptonic decays of $\W$ bosons in events with lower $\MET$; this
search has more sensitivity than the first as $m_{\chiz_1}$ approaches
$\mchi$.

\subsection{Searches with three leptons using \texorpdfstring{$\MET$}{MET} shape}
\label{trilepton-broad}

For our study of three-lepton events with significant $\MET$, we make
use of our previous analysis~\cite{SUS-11-013-paper}, based on the
same data sample as the present study.  The analysis requires three
leptons (only electrons or muons) and $\HT<200\GeV$, where $\HT$ is
the scalar sum of the $\PT$ of the jets in the event.  OSSF dileptons
are rejected if $75\GeV<\mdil<105\GeV$ in order to suppress background
from $\Z$ bosons.  For the lepton selection, at least one electron or
muon is required with $\PT>20\GeV$, and another with $\PT > 10\GeV$;
the third lepton must have $\PT > 8\GeV$; this search additionally
requires $|\eta| < 2.1$ for all three leptons. A more detailed
description of the analysis can be found in
Ref.~\citen{SUS-11-013-paper}.

The number of events observed for $\MET>50\GeV$ and the corresponding
background predictions are given in Table~\ref{tab-sus-013-fig3} in
10-\GeVns-wide bins (corresponding to the display of the same data in
Fig.~3 (left) of Ref.~\citen{SUS-11-013-paper}).  The analysis in
Ref.~\citen{SUS-11-013-paper} considers two regions of \MET only:
$\MET < 50\GeV$ and $\MET > 50\GeV$. In the present study, we take
this latter region and use the separate contents of the bins in
Table~\ref{tab-sus-013-fig3} in a combined statistical treatment. This
approach provides more powerful discrimination between signal and
background than the treatment of Ref.~\citen{SUS-11-013-paper},
because of the different shapes of signal and background across these
bins.

All details of the event selection, background estimates, and
evaluation of systematic uncertainties are as described in
Section~\ref{detector} and Ref.~\citen{SUS-11-013-paper}.  Briefly,
efficiencies of electron/muon identification and isolation
requirements are estimated using the method described in
Ref.~\cite{WZpaper} for Z\,$\to\ell^+\ell^-$ events, and are in
agreement with the simulation to within 2\% ($1$\%) for electrons
(muons). Background due to Drell-Yan processes (including $\zjets$
boson production), with a jet providing a third genuine (non-prompt)
lepton or a hadron misidentified as a lepton, is evaluated from studies
of isolated tracks failing or passing electron/muon identification
criteria, separately for samples enriched in heavy- and light-flavor
jets. This background decreases rapidly to negligible levels for
$\MET>50\GeV$.  The main backgrounds for $\MET>50\GeV$ are from
diboson and $\ttbar$ production and are estimated from the simulation.

\begin{table}[h]
   \begin{center}
\topcaption{The observed and mean expected background in bins of \MET
for three-lepton events with $\HT<200\GeV$, an opposite-sign
same-flavor (OSSF) lepton pair, and no \Z boson candidate.  These
results correspond to the distributions shown in Fig.~3 (left) of
Ref.~\citen{SUS-11-013-paper}.  Uncertainties include statistical and
systematic contributions.
\label{tab-sus-013-fig3}
}
   \begin{tabular}{@{} lcc @{}} % Column formatting, @{} suppresses leading/trailing space
      \hline\hline
      \MET Range (GeV) & Observation & Background \\
      \hline
50-60 & 5 &   7.01  $\pm$  2.15\\
60-70 & 10 &  5.36  $\pm$  1.46\\
70-80 & 2 &   3.35  $\pm$  0.93\\
80-90 & 5 &   2.52  $\pm$  0.68\\
90-100 & 1 &  2.14  $\pm$  0.56\\
100-110 & 0 & 2.37  $\pm$  0.83\\
110-120 & 3 & 1.49  $\pm$  0.47\\
120-130 & 1 & 1.06  $\pm$  0.32\\
130-140 & 0 & 0.38  $\pm$  0.11\\
140-150 & 2 & 0.26  $\pm$  0.10\\
150-160 & 0 & 0.15  $\pm$  0.06\\
160-170 & 1 & 0.16  $\pm$  0.06\\
170-180 & 0 & 0.08  $\pm$  0.03\\
180-190 & 0 & 0.54  $\pm$  0.42\\
190-200 & 0 & 0.05  $\pm$  0.03\\
$>$200 & 0 & 0.33  $\pm$  0.16\\
      \hline\hline
\end{tabular}
\end{center}
\end{table}

Section~\ref{conclusion} presents the detailed interpretation of these
results.

\subsection{Searches with three leptons using \texorpdfstring{$\mdil$
and $\MT$}{dilepton mass and transverse mass}}
\label{trilepton-targeted}

The alternative three-lepton search, based on $\mdil$ and $\MT$,
introduces in addition a veto on events having an identified
\bjetnohyphen (using the TCHE medium working point) with $\PT>20\GeV$.
By vetoing only \bjetsnohyphen, this requirement suppresses $\ttbar$
background while avoiding exposure to signal loss (for example due to
initial-state radiation) from a more general jet veto.

We require at least one electron or muon with $\pt > 20\GeV$ and two
more with $\pt > 10\GeV$, all with $|\eta|<2.4$.  After requiring
$\MET > 50\GeV$ (and making no requirement on $\HT$), events are
characterized by the values of the invariant mass $\mdil$ of the OSSF
pair, and the transverse mass \MT formed from the \MET vector and the
transverse momentum $\PT^{\ell}$ of the remaining lepton:
\begin{equation}
       \MT \equiv \sqrt{2\MET p_{\mathrm{T}}^{\ell}[1-\cos(\Delta \phi_{\ell,\MET})]}.
\end{equation}

For three-muon and three-electron events, the OSSF pair with $\mdil$
closer to the $\Z$ mass is used.  For backgrounds where a true OSSF
pair arises from a low-mass virtual photon, this can result in a
misassignment; simulation of this effect is validated with identified
$\mu\mu\Pe$ and $\mu\Pe\Pe$ events by treating all three leptons as
having the same flavor.

\subsubsection{Background due to $\W\Z$ production}
\label{sec:WZ}

The largest background is due to SM $\W\Z$ production in which both
bosons decay leptonically.  Studies with data indicate that the
simulation-based estimates of systematic uncertainties on both the
$\W\Z$ background characteristics and signal resolutions are generally
reliable, but especially at high-\MT, corrections are obtainable
through detailed comparisons of data and the simulation.  Here, we
present one such study: the calibration of the hadronic recoil of the
$\W\Z$ system.  In addition, the overall $\W\Z$ event yield
normalization is validated using events where $\mdil$ and $\MT$ are
consistent with the \Z and \W boson masses ($81\GeV<\mdil<101\GeV$,
$\MT<100$\GeV), respectively.  We find good agreement with the SM
simulations, as presented below.

The simulation of \MET (and hence $\MT$) is corrected using a
generalization of the \Z-recoil method used in the CMS measurements of
the $\W$ and $\Z$ cross sections~\cite{WZpaper}.  The transverse
hadronic recoil vector $\vec{u}_{\mathrm{T}}$ is
\begin{equation}
\vec{u}_{\mathrm{T}}= -\vecMET-\vec{p}_{{\mathrm{T}},1}-\vec{p}_{{\mathrm{T}},2}
\end{equation}
for \Z events and
\begin{equation}
\vec{u}_{\mathrm{T}}= -\vecMET-\vec{p}_{{\mathrm{T}},1}-\vec{p}_{{\mathrm{T}},2}-\vec{p}_{{\mathrm{T}},3}
\end{equation}
for $\W\Z$ events, where $\vecMET$ is the missing transverse energy
vector, and $\vec{p}_{{\mathrm{T}},i}$ is the transverse momentum
vector of each of the two leptons from the \Z decay or three leptons
from the $\W\Z$ decay.  The recoil vector is resolved into components:
$u_1$ parallel to the direction of the respective \Z or $\W\Z$ system,
and $u_2$ perpendicular to the \Z or $\W\Z$ direction (known in the
simulation and approximated in the data).  The $u_1$ component is
sensitive to calorimeter response and resolution, while the $u_2$
component is predominantly determined by the underlying event and
multiple interactions.  Using a pure sample of \Z boson events,
detailed studies of both components as a function of the \Z boson
$\PT$ value yield corrections to the simulation, which are implemented
event-by-event assuming that the results for Z production are similar
to those for $\W\Z$ production.  These data-based corrections alter
the expected background by up to 25\%, and allow us to reduce the
systematic uncertainty associated with the simulation.

Reconstructed leptonic decays of $\Z$ bosons are used to calibrate
lepton energy scales and resolutions, separately for electrons and
muons, in bins of $\PT$ and $\eta$.  The uncertainties from this
procedure are propagated into uncertainties on the mean background
estimation by using the simulation.  Table \ref{tab:WZsyst} summarizes
these and the other systematic uncertainties in the estimation of the
$\W\Z$ background.

\begin{table}[!htbp]
\topcaption{Relative systematic uncertainties for the mean $\W\Z$
background.  ``On-\Z'' refers to events in which the OSSF pair
satisfies $81<\mdil<101\GeV$.  ``Off-\Z'' refers to events with either
$\mdil<81\GeV$ or $\mdil>101\GeV$. The events are further categorized
according to whether they have low ($<100\GeV$) or high ($>100\GeV$)
\MT values.  The ``Off-\Z, low-\MT'' column corresponds to the sum of events in
regions I and V in Fig.~\ref{fig:DataYields}, while the ``Off-\Z, high
\MT'' column corresponds to the sum of regions II and IV.
}
\begin{center}
\begin{tabular}{l|ccc}
\hline\hline
     & On-$\Z$, high-\MT  & Off-Z, low-\MT  & Off-$\Z$, high-\MT  \\ \hline
Hadronic recoil & 29.7\% & 0.9\% & 14.9\% \\
$\W\Z$ versus $\Z$ recoil & 7.2\% & 0.5\% & 3.4 \%  \\
Lepton energy scale & 1.8\% & 0.7\% & 0.7\%  \\
Lepton energy resolution & 1.4\% & 6.9\% & 4.5\%  \\
Boson $\pt$ & 5.1\% & 0.4\% & 2.2\%  \\
$\Z$ mass shape & 0.2\% & 0.4\% & 2.5\% \\
Normalization &  9.3\% & 9.3\% & 9.3\%    \\
\hline
\hline
Sum & 32.4\% & 11.7\% & 18.8\% \\
\hline\hline
\end{tabular}
\label{tab:WZsyst}
\end{center}
\end{table}

\subsubsection{Background due to $\ttbar$ production and other processes}
\label{sec-ttbar}
The second-largest background is from events with two genuine isolated
prompt leptons and a third identified lepton that is either a
non-prompt genuine lepton from a heavy-flavor decay or a misidentified
hadron, typically from a light-flavor jet.  Top-quark pair, \zjets,
and \wwjets events are the main processes that contribute to this
background. We measure this background using control samples in
data. The probability for a non-prompt lepton to satisfy the isolation
requirement ($\Irel < 0.15$) is measured in a data sample enriched
with QCD dijet events, and varies from 2\% to 3\% for muons and from
6\% to 8\% for electrons as a function of lepton \PT. These
probabilities, applied to the three-lepton events where the isolation
requirement on one of the leptons is removed, are used to estimate
background due to such non-prompt leptons.

Another background studied with data is the rare process in which a \Z
boson is accompanied by an initial- or final-state radiation photon
that converts internally or externally, leading to a reconstructed
three-lepton final state when the conversion is highly asymmetric
\cite{SUS-11-013-paper}.

The systematic uncertainties assigned to the $\ttbar$ background and
other backgrounds studied with data are based on differences between
the predicted and true yields when the method is applied to simulated
events, as well as on the effect of the prompt-lepton contamination in
control samples.

Backgrounds from very rare SM processes that have not yet been
adequately measured in the data ($\ZZ$, $\ttbar\Z$, $\ttbar\W$,
three-vector-boson events) are estimated from simulation.  For these
sources, a systematic uncertainty of 50\% is assigned to account for
uncertainty in the NLO calculations of cross sections.

\subsubsection{Observations in the three-lepton search with $\mdil$ and \MT}
\label{sec:FinalResults}

\begin{figure}[b!]
\begin{center}
\includegraphics[scale=0.60]{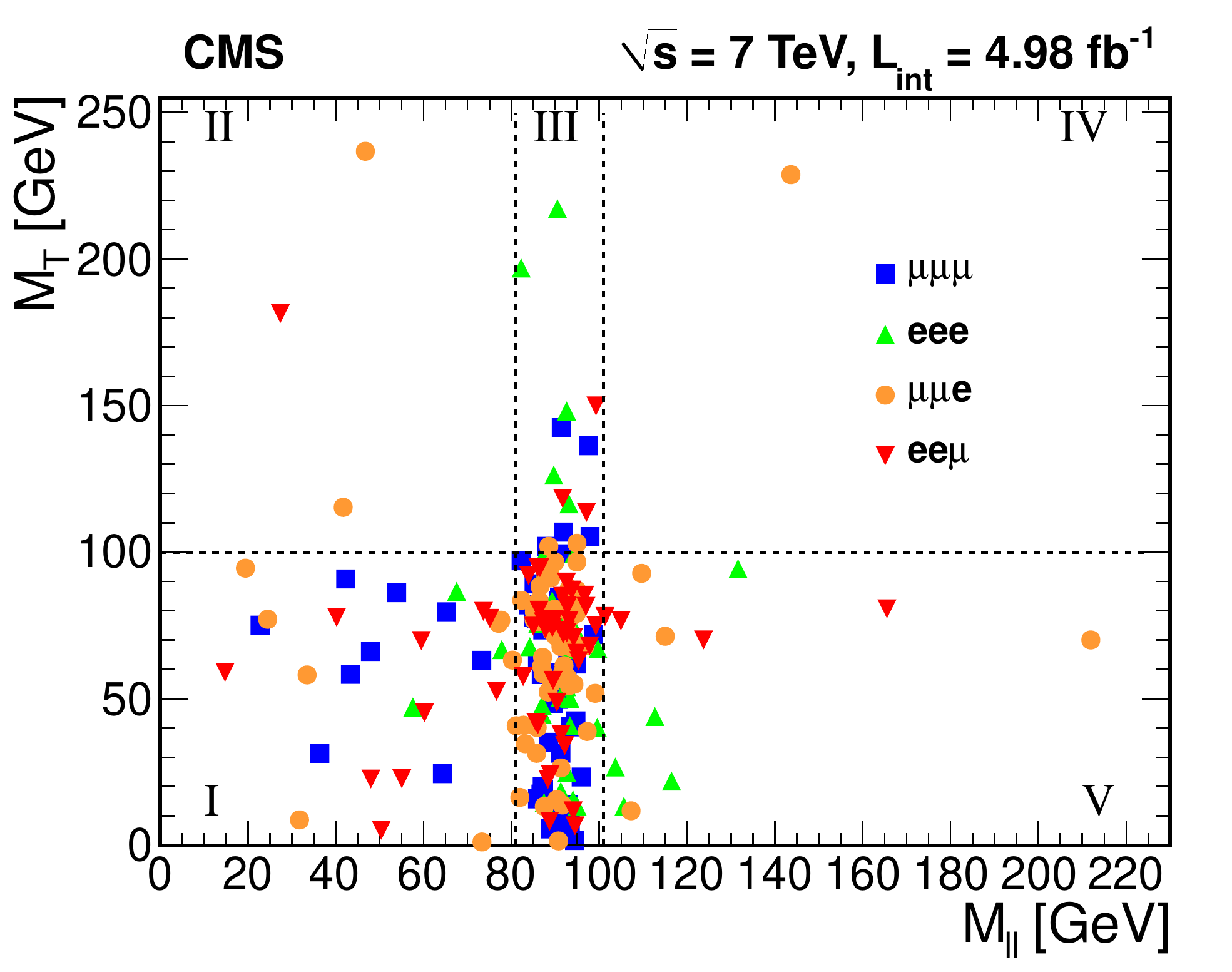} \\
\caption{$\MT$ versus $\mdil$
for the selected events in data. (Unlabeled Region VI lies between
Regions I and V.)  Two events appear outside the limits of the plot;
one is a $\mu\mu\mu$ event at ($\mdil,\MT$) = ($240\GeV,399\GeV$) and
the other is an $\Pe\Pe\Pe$ event at ($95\GeV,376\GeV$).
\label{fig:DataYields}
}
\end{center}
\end{figure}

Figure~\ref{fig:DataYields} presents a scatter plot of $\MT$ versus
$\mdil$ for the selected events.  The dashed lines divide the plane
into six regions. The horizontal dashed line at $\MT=100\GeV$
separates the lower-$\MT$ region, which contains most of the
background associated with on-shell $\W$ bosons, from the region
depleted of this background.  The vertical dashed lines at
$\mdil=81\GeV$ and $101\GeV$ define the endpoints of the region
dominated by \Z boson decays.  In the lower $\mdil$ region, the search
is sensitive to the signal production process of
Fig.~\ref{fig:charginos-slep} with small to moderate
$\chiz_2$--$\chiz_1$ mass splittings ($<100\GeV$), while being subject
to background from $W +\gamma^{*}/\Z^*$ events, especially in Region
I. In the higher-$\mdil$ region, the search is sensitive to models
with larger mass splittings. Region VI (on-Z, low $\MT$) is dominated
by $\W\Z$ and $\Z\Z$ backgrounds. Leakage from this region
contaminates the nearby regions.

\begin{figure}[t!]
\begin{center}
\begin{tabular}{ccc}
\subfigure[]{\includegraphics[scale=0.4]{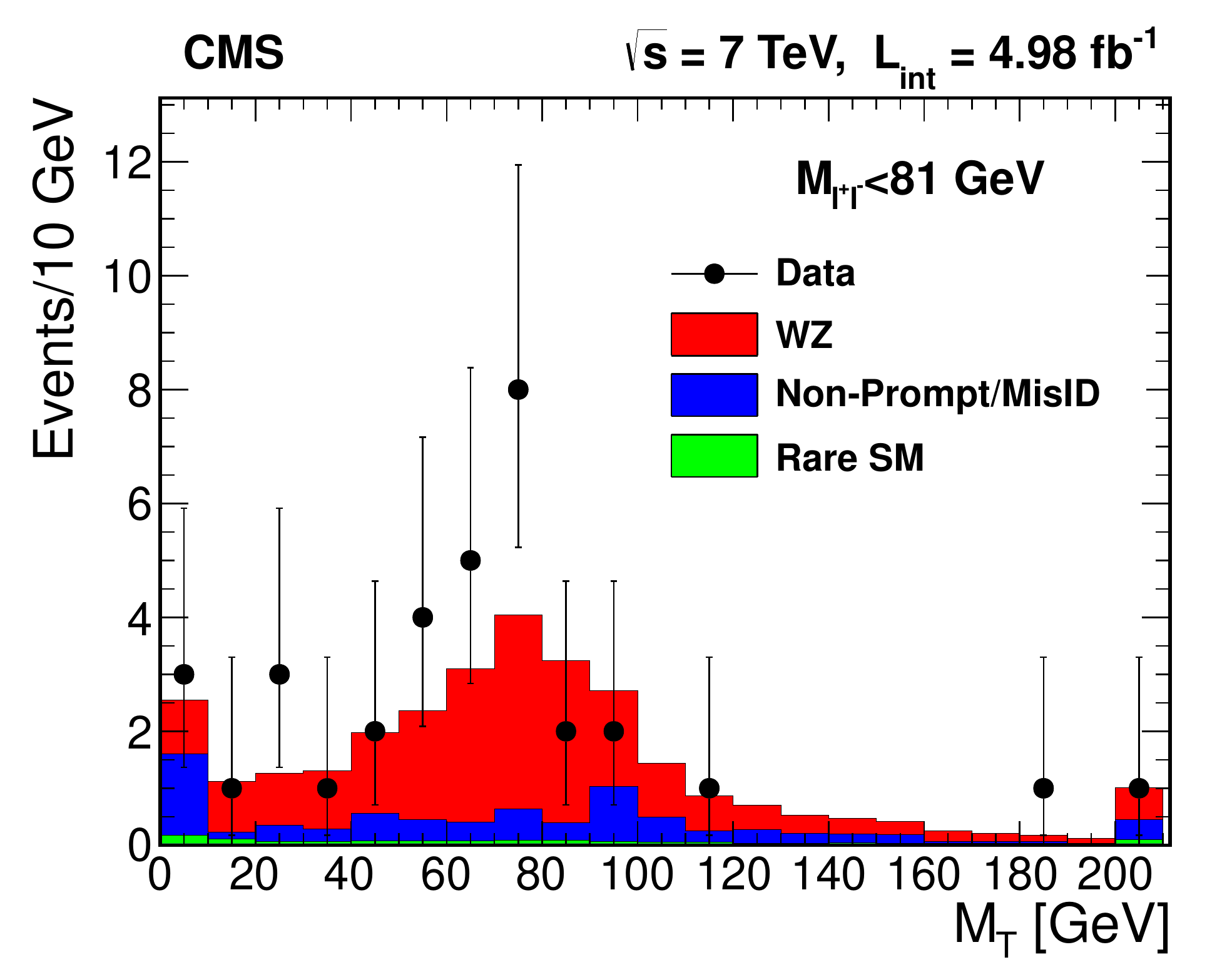} }
\subfigure[]{\includegraphics[scale=0.4]{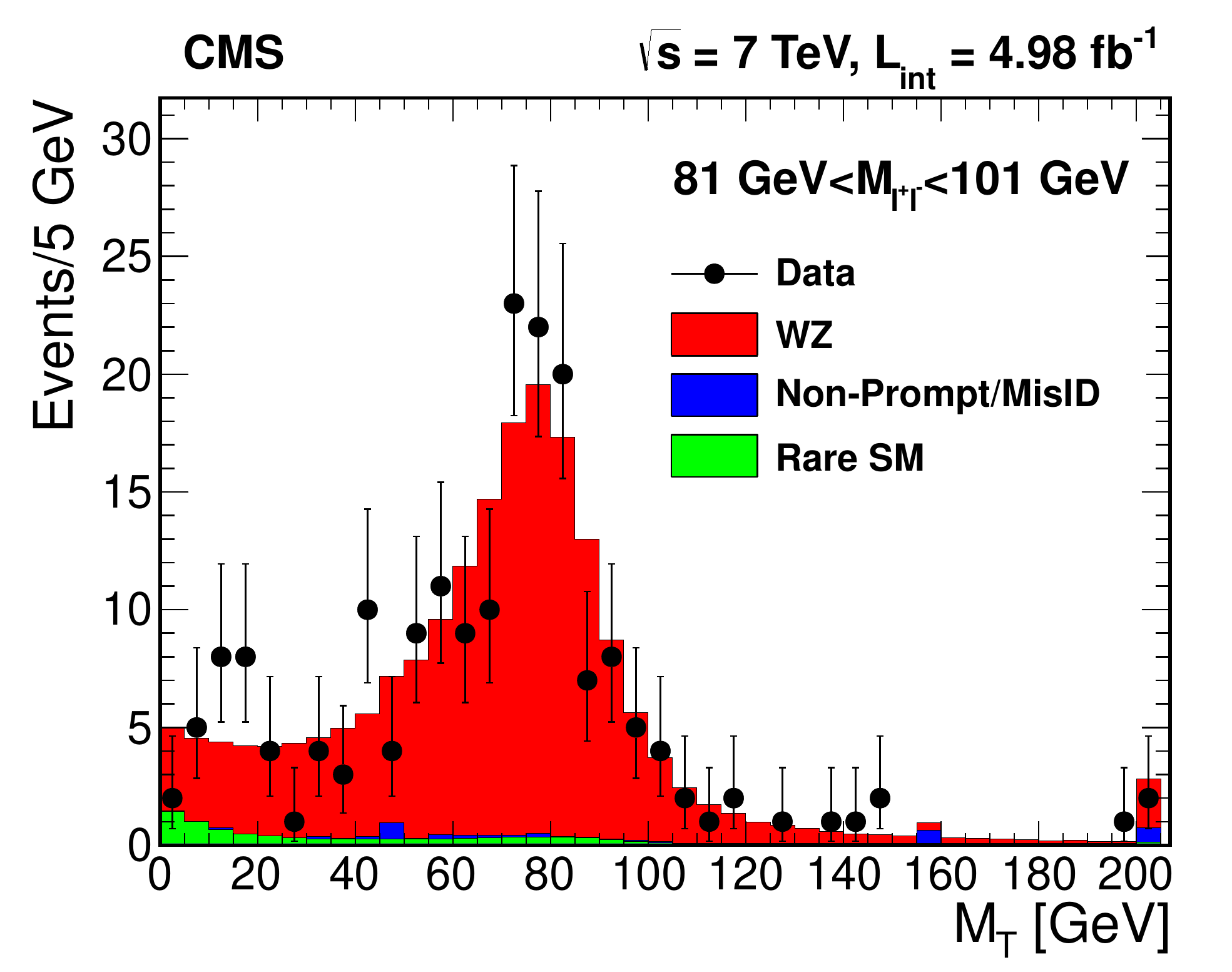} }\\
\subfigure[]{\includegraphics[scale=0.4]{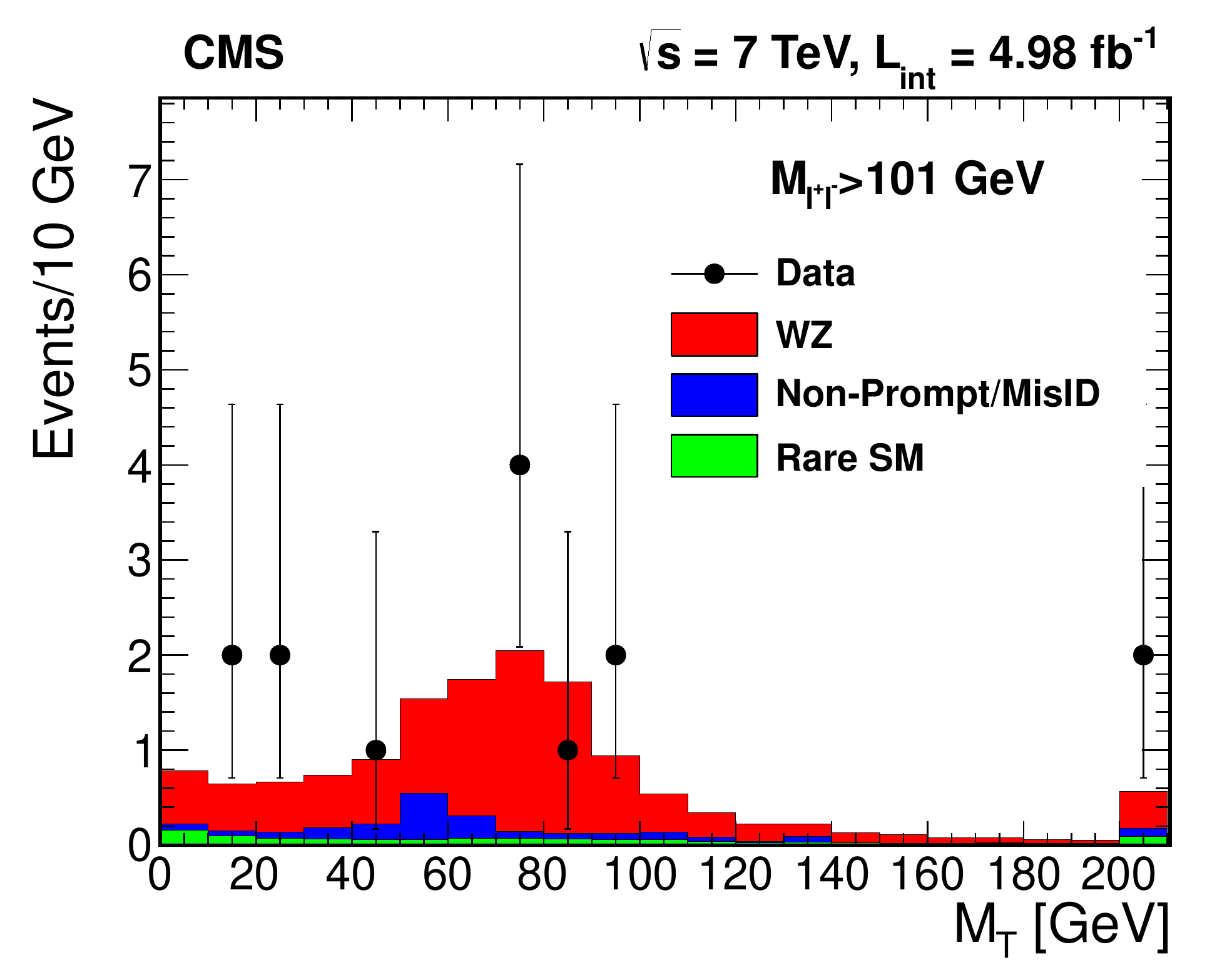} }
\end{tabular}
\topcaption{Observed and mean expected \MT distribution for $\mdil$
in the regions (a) below the \Z boson mass, (b) on the \Z boson mass,
and (c) above the \Z boson mass. Rare SM processes include
three-vector-boson production, production of top-quark pairs together
with a vector boson, and \ZZ production. The last bin in each
histogram includes the events with $\MT$ beyond the histogram range.
\label{fig:mTcomparison}
}
\end{center}
\end{figure}

Figure~\ref{fig:mTcomparison} shows the $\MT$ distributions for data
and the mean expected SM background below the $\Z$ (Regions I and II),
on-$\Z$ (Regions III and VI), and above the $\Z$ (Regions IV and V).
The background shape from non-prompt or misidentified leptons is taken
from simulation while the normalization is derived from the data.

Table~\ref{tab:limits} contains a summary of the mean estimated
backgrounds and observed yields.  There is no evidence for a signal,
and the background shape is well reproduced within the limited
statistics.

\begin{table}[!ht]
\begin{center}
\topcaption{Summary of mean expected backgrounds and observations in
each region for the three-lepton search based on $\mdil$ and \MET.
Uncertainties include statistical and systematic contributions.
\label{tab:limits}
}
\begin{tabular}{c c c c c c } \hline\hline
Region & $\W\Z$ & Non-prompt & Rare SM & Total background & Data  \\ \hline
I    & 16.2 $\pm$ 2.9 & 4.7 $\pm$ 2.4 & 2.1 $\pm$ 1.5 & 23.0 $\pm$ 5.1 & 31     \\
II   & 3.6 $\pm$ 0.8 & 1.94 $\pm$ 1.02 & 0.4 $\pm$ 0.2 & 6.0  $\pm$ 1.3 & 3      \\
III  & 15.6 $\pm$ 5.7 & 0.2 $\pm$ 0.1 & 0.8 $\pm$ 0.4  & 16.6 $\pm$ 5.7 & 17     \\
IV   & 1.6 $\pm$ 0.4 & 0.2 $\pm$ 0.1 & 0.4 $\pm$ 0.2 & 2.2  $\pm$ 0.5 & 2      \\
V    & 8.7 $\pm$ 1.7 & 1.4 $\pm$ 0.8 & 0.9 $\pm$ 0.4  & 11.0 $\pm$ 1.9 & 12     \\
VI   & 150.6 $\pm$ 25.7 & 2.6 $\pm$ 1.4 & 11.7 $\pm$ 5.8 & 164.9 $\pm$ 26.4 & 173   \\
\hline\hline
\end{tabular}
\end{center}
\end{table}

Section~\ref{conclusion} contains the detailed interpretation of these
observations, which are found to have comparable sensitivity to the
$\MET$-based search of Section~\ref{trilepton-broad}.

\section{Searches in the same-sign two-lepton final state}
\label{dilepton}
Three-lepton final states are not sensitive to direct
chargino-neutralino production if one of the leptons is unidentified,
not isolated, or outside the acceptance of the analysis.  The CMS
detector has high geometrical acceptance for all leptons. However,
when the signal-model mass splittings are such that one lepton has
$\PT<10\GeV$, three leptons are unlikely to be selected.  Some of
these otherwise-rejected events can be recovered by requiring only two
leptons, which should however be of same sign (SS) to suppress the
overwhelming background from opposite-sign
dileptons~\cite{Nachtman:1999ua,Matchev:1999nb}.

The SS dilepton search requires at least one electron or muon with
$\pt > 20\GeV$, and another with $\pt > 10\GeV$, with $|\eta|<2.4$ for
both.  We exclude events that contain a third lepton, using the
criteria of Section~\ref{trilepton-targeted}, in order to facilitate
combination with those results.  Furthermore, as events with $\tau$
leptons can be important in some SUSY scenarios~\cite{Lykken:1999kp},
we include the $\Pe\tau$, $\mu\tau$, and $\tau\tau$ final states; for
this purpose, we use hadronic decays of the $\tau$.  The isolation
criteria for hadronically decaying $\tau$ leptons require that, apart
from the hadronic decay products, there be no charged hadron or photon
with $\PT$ above $0.8\GeV$ within a cone of $\Delta R = 0.5$ around
the direction of the $\tau$.

An important class of background for SS events is that with one
genuine prompt lepton and either a non-prompt genuine lepton from a
heavy-flavor decay or a misidentified hadron.  This background arises
mainly from events with jets and a $\W$ or $\Z$ boson. Much of the
analysis strategy is driven by the need to suppress these events.
Electron and muon selection criteria are thus tightened: the isolation
criterion becomes $\Irel<0.1$, and we add a criterion to limit the
maximum energy deposit of muon candidates in the calorimeters.

Events containing OSSF pairs with $|\mdil - M_Z|<15\GeV$ are
eliminated in order to reduce background due to processes such as
$\W\Z$ and $\ttbar\Z$ production.  For this purpose we
select these events by using
looser isolation criteria ($\Irel<1.0$ for muons and barrel electrons,
and $\Irel<0.6$ for endcap electrons) and looser identification
requirements for electrons.

The remaining background with a non-prompt lepton is estimated with
techniques described in Ref.~\citen{SUS-11-010}, where the probability
for a non-prompt lepton to pass the signal selection is derived from
control regions in data using extrapolations in the isolation and
identification criteria. The systematic uncertainty on these
predictions is 50\% for light leptons and 30\% for $\tau$ leptons.

Residual background is mostly due to charge misassignment in events
with an OSSF pair, e.g., from Drell-Yan, $\ttbar$, or $\W\W$
processes.  We quantify the charge misassignment probability for
electrons and $\tau$ leptons by studying SS $\Pe\Pe$ or $\tau\tau$
events inside the $\Z$ mass peak region in data. For electrons, this
probability is $0.0002 \pm 0.0001$ in the ECAL barrel and $0.0028 \pm
0.0004$ in the ECAL endcap, and for $\tau$ leptons it is $0.009\pm
0.024$. For muons, it is determined from cosmic-ray data to be of
order $10^{-5}$ and is neglected.

Backgrounds of lesser importance include those from rare SM processes
such as diboson production, associated production of a $\ttbar$ pair
with a vector boson, or triboson production.  They are taken from
simulation with a 50\% systematic uncertainty assigned. An exception
is $\W\Z$ production, for which normalization to the measured cross
section is available, thus reducing the systematic uncertainty to
20\%.

\begin{figure}[t!]
\begin{center}
\subfigure[]{\includegraphics[scale=0.4]{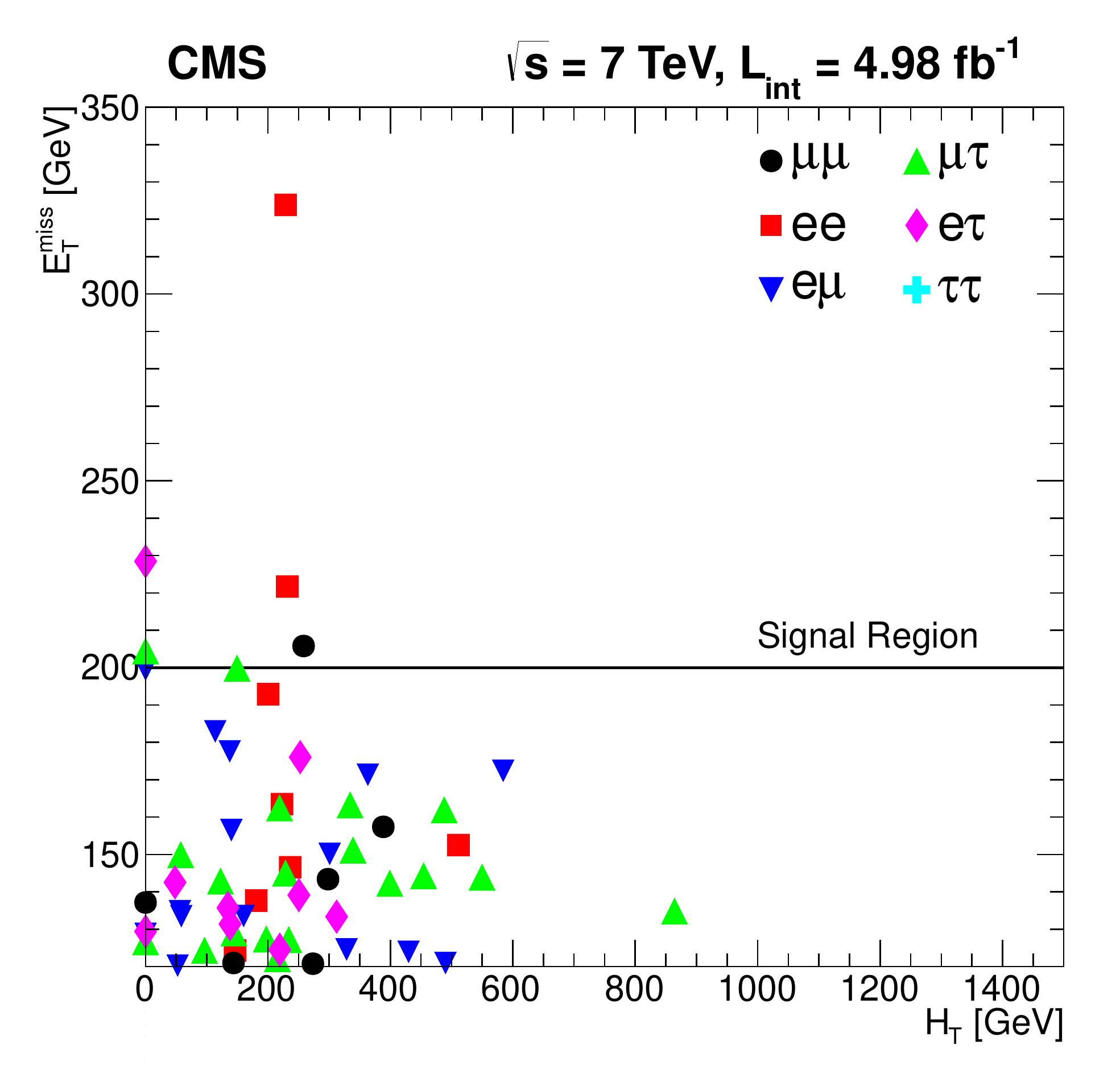} }
\subfigure[]{\includegraphics[scale=0.4]{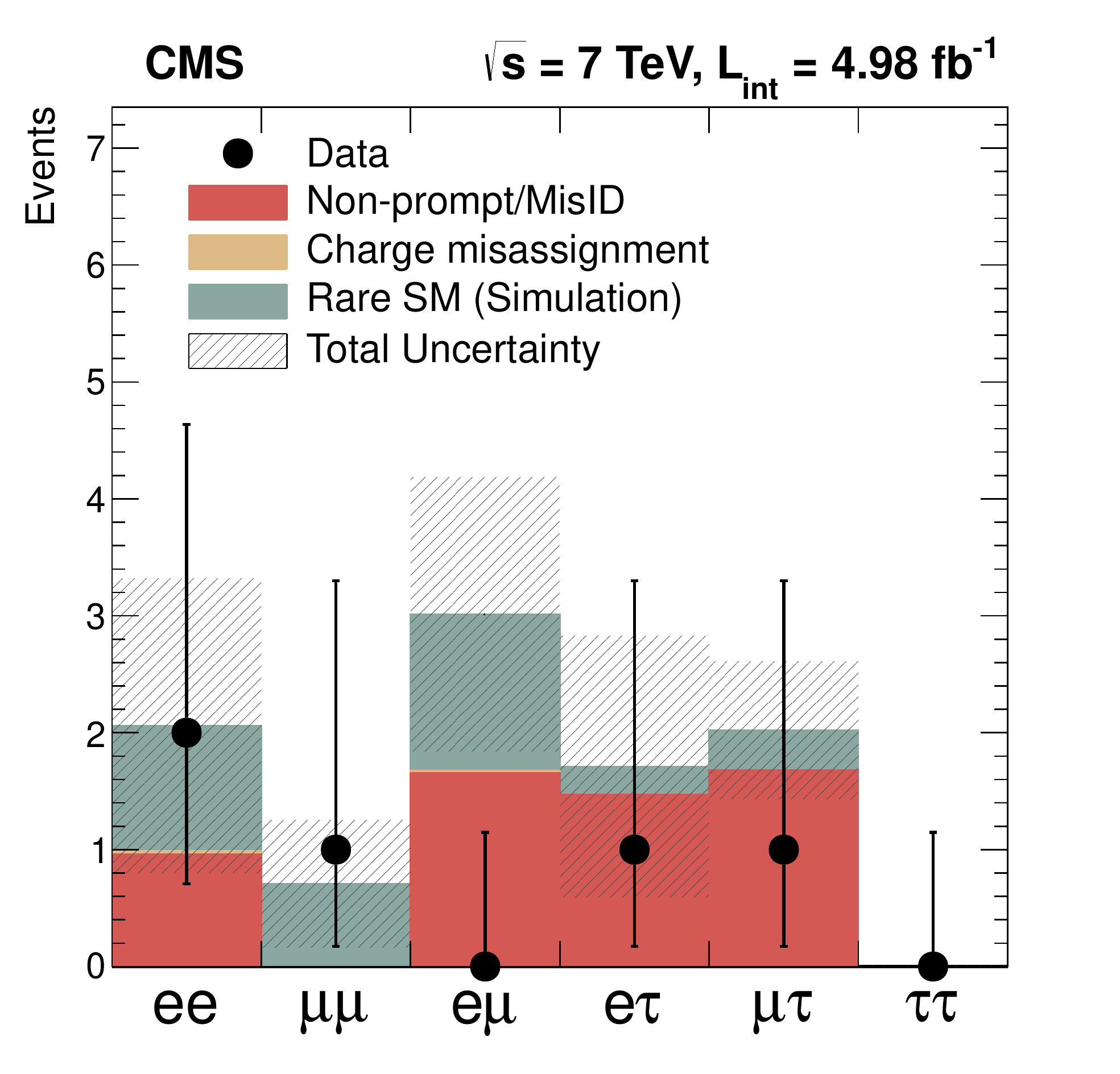} }
\caption{(a) \MET versus \HT for same-sign dilepton candidate events.
(b) Mean expected background yields with their uncertainty and
observed number of events in the six channels, for the signal region
(\MET $> 200 \GeV$).
\label{fig:combinedtau}
}
\end{center}
\end{figure}

The distribution of events thus selected is studied in the plane of
\MET versus \HT, as displayed in Fig.~\ref{fig:combinedtau}(a).  The
signal region is defined by the criterion $\MET>200\GeV$, with the
$120\GeV < \MET < 200\GeV$ interval used as a control region to
confirm understanding of backgrounds.  In the control region, the
total mean expected background for events without a $\tau$ ($\Pe\Pe$,
$\mu\mu$, and $\Pe\mu$ events) is 24.8 $\pm$ 7.6, and 27 events are
observed.  The total mean expected background for $\Pe\tau$,
$\mu\tau$, and $\tau\tau$ events is 24.5 $\pm$ 8.9, and 26 events are
observed. The observed signal region yields in the various
lepton-flavor final states are displayed in
Fig.~\ref{fig:combinedtau}(b).  Table~\ref{tab:yields3rdveto} presents
the mean expected background and the observed yields in the signal
region.  Section~\ref{conclusion} presents the detailed interpretation
of these observations; combining the same-sign dilepton search with
the three-lepton search increases the mass limits by up to
approximately $20 \GeV$.

\nobreak  %to try to tune placement
\begin{table}%[b!]
	\begin{center}
\topcaption{Summary of mean expected backgrounds and observed yields in
the \MET $> 200 \GeV$ signal region for all six same-sign dilepton
channels.  The background categories comprise non-prompt and
misidentified leptons, charge misassignment, and rare SM processes.
Uncertainties include statistical and systematic contributions.
\label{tab:yields3rdveto}
}
		\small
		\begin{tabular}{lccccccc}
			\hline \hline
Source			                &     $\Pe\Pe$         &     $\mu\mu$    &     $\Pe\mu$      &     $\Pe\tau$    &   $\mu\tau$     &   $\tau\tau$     &    Sum     \\ \hline
			Non-pr/misID           &  1.0 $\pm$ 0.8 &  0.0 $\pm$ 0.2 &  1.7 $\pm$ 1.0  &  1.5 $\pm$ 1.1 &  1.7  $\pm$ 0.6 &  0.00 $\pm$ 0.00 &  5.8 $\pm$ 1.9  \\
			Charge misass    &  0.0 $\pm$ 0.0 &        --        &  0.0 $\pm$ 0.0  &  0.0 $\pm$ 0.0 &  0.0  $\pm$ 0.1 &  0.00 $\pm$ 0.01 &  0.1 $\pm$ 0.1  \\
			Rare SM         &  1.0 $\pm$ 0.7 &   0.7 $\pm$ 0.5 &  1.3 $\pm$ 0.7  &  0.3 $\pm$ 0.1 &  0.4  $\pm$ 0.2 &  0.00 $\pm$ 0.00 &  3.7 $\pm$ 1.5  \\ \hline \hline
			Total background       &  2.1 $\pm$ 1.0 &   0.7 $\pm$ 0.5 &  3.1 $\pm$ 1.2  &  1.7 $\pm$ 1.1 &  2.0  $\pm$ 0.6 &  0.00 $\pm$ 0.01 &  9.5 $\pm$ 2.4  \\ \hline \hline
			Observed        &  2             &   1             &  0              &  1             &  1              &  0               &  5              \\ \hline \hline
			\end{tabular}
	\end{center}
\end{table}

The same-sign analysis is potentially sensitive to the processes of
Fig.~\ref{fig:charginos-slep} in the $\tau$-dominated scenario, in
which the chargino and neutralino both decay only to a $\tau$.  With
the present selection, we are only able to exclude a limited region of
phase space for this scenario, bounded by $m_{\chiz_1} < 50\GeV$ and
$\mchi < 250\GeV$.

\section{Searches in the
\texorpdfstring{$\W\Z/\Z\Z + \MET$}{WZ/ZZ + MET} final state with two
leptons and two jets}
\label{diboson}
Finally, we consider events with two on-shell vector bosons and
significant \MET{}.  Ref.~\citen{SUS-11-013-paper} presents results
relevant for the four-lepton final state, corresponding to the
two-\Z{}-boson process of Fig.~\ref{fig:charginos-wz}(b), when each \Z
boson decays either to an electron or a muon pair.  In the following,
we extend sensitivity to both diagrams of Fig.~\ref{fig:charginos-wz}
by selecting events in which a $\Z$ boson decays to either $\Pe\Pe$ or
$\mu\mu$, while a $\W$ boson or another $\Z$ boson decays to two jets. SM
diboson events with the corresponding final states do not contain
intrinsic \MET.

This search is an extension of our previous
result~\cite{SUS-11-021-paper}.  We use the same selection of jets,
leptons, and \MET, as well as the same background estimation methods.
Both leptons must have $\PT> 20\GeV$.  In particular, jets are
required to have $\PT>30\GeV$ and $|\eta|<3$.
The \MET signal regions are indicated in Table \ref{resulttableex},
with the entries indicating mean background estimates after applying
all selection criteria described below.

We suppress background from $\ttbar$ events by a factor of approximately 10
by rejecting events that contain an identified \bjetnohyphen.  We use
the TCHE loose (medium) working point for jets with $\PT<100\GeV$
($>100\GeV$).  Further suppression of the $\ttbar$ and $\zjets$
background is achieved by requiring that the dijet mass $\mjj$ be
consistent with a $\W$ or $\Z$ boson, namely $70\GeV < \mjj <
110\GeV$.  Background from $\wzjets$ events is suppressed by rejecting
events that contain a third identified lepton with $\PT>20\GeV$.

Background from SM $\zjets$ events with artificial \MET from jet
mis-measurements must be carefully estimated, since the artificial
\MET is not necessarily well-reproduced in simulation.  Using the
method described in Ref~\cite{SUS-11-021-paper}, a control sample of
$\gjets$ events is used to model the \MET in $\zjets$ events, after
performing a reweighting procedure to take into account the different
kinematic properties of the hadronic systems in the control and signal
samples.

Background processes with uncorrelated flavor, while dominated by
$\ttbar$ events, also include events with $\tau\tau$ (via Drell-Yan production and
followed by leptonic decays), $\W\W$, and single top production.  For these
processes, production in the same-flavor $\Pe\Pe$ and $\mu\mu$ final
states used for the search is modeled using a control sample of
opposite-flavor (OF) $\Pe\mu$ events. Subdominant background
contributions from SM $\W\Z$ and $\Z\Z$ production are estimated from
simulation.

\begin{figure} [!ht]
  \begin{center}
	\includegraphics[scale=0.50]{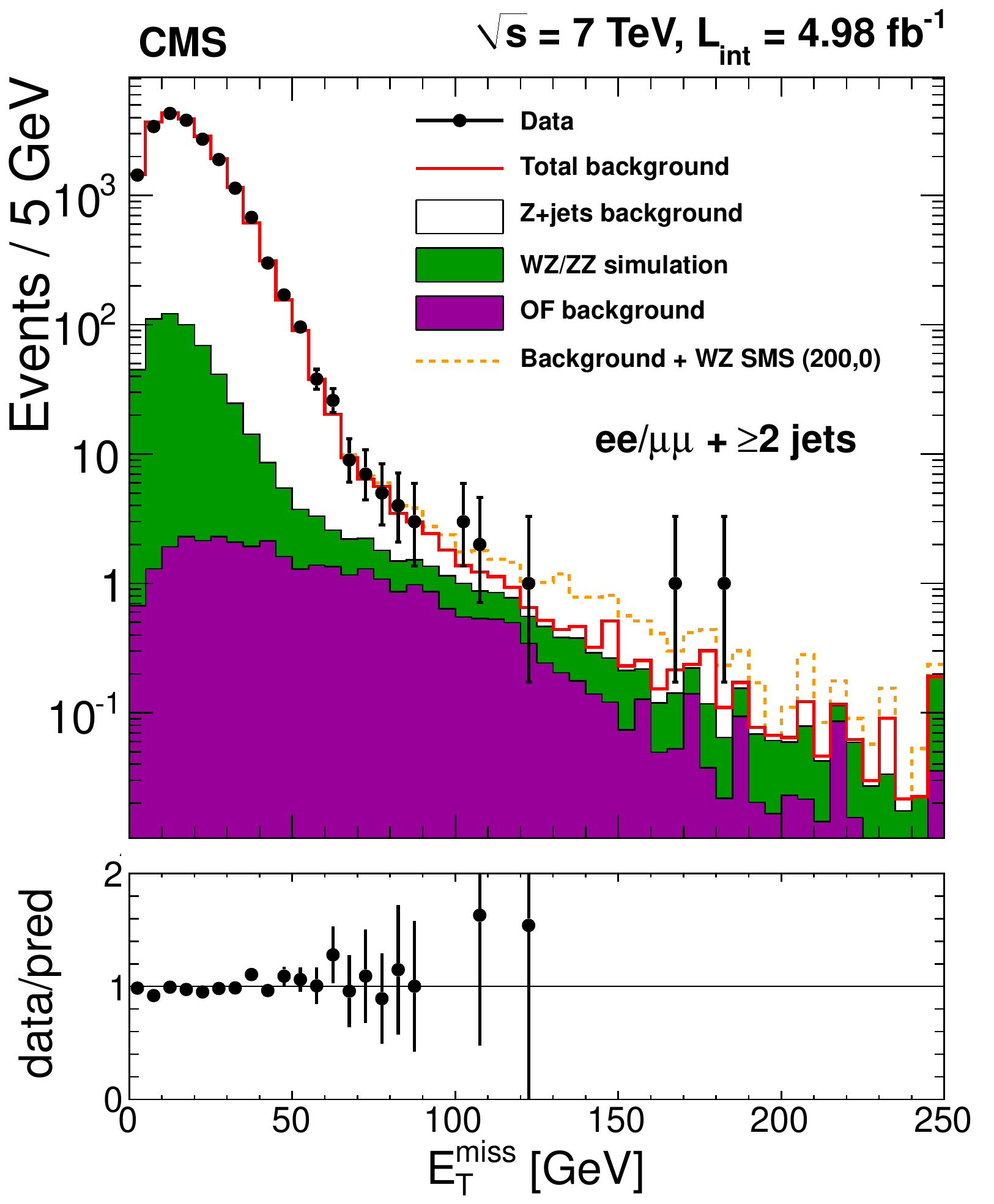}
	\caption{
\label{fig:pfmet_eemm}\protect
Observed \MET distribution for \wzmet events after all selection
criteria are applied except that on \MET (solid points), in comparison
with the corresponding SM expectation.
For purposes of illustration, the \MET distribution
expected for $\W\Z$ SMS events with $\mchi = 200 \GeV$ and a massless
LSP is shown.  The plot below the main figure shows the ratio of the
observed and mean-expected-background distributions.}
\end{center}
\end{figure}

The mean expected backgrounds in bins of \MET and the observed yields
are summarized in Table \ref{resulttableex} and displayed in
Fig.~\ref{fig:pfmet_eemm}.  Section~\ref{conclusion} contains the
interpretation of these results, including a combination with those of
Ref.~\citen{SUS-11-013-paper}.

\begin{table}[ht]
  \begin{center}
\topcaption{\label{resulttableex}
Summary of mean expected backgrounds and observed data in each of the
\MET signal regions, in final states with two opposite-sign leptons,
two jets, and \MET.  The total background is the sum of the \zjets
background evaluated with \gjets events, the flavor-symmetric
background evaluated from opposite-flavor events (OF background), and
the $\W\Z$/$\Z\Z$ background expected from simulation ($\W\Z$/$\Z\Z$
background).
Uncertainties include statistical and systematic contributions.
}
\begin{tabular}{lccc}
\hline\hline
Source & $30 \le \MET < 60\GeV$ & $60 \le \MET < 80\GeV$ & $80 \le \MET < 100\GeV$ \\
\hline

\zjets background &  2298 $\pm$  737  &    32.9 $\pm$   11.1  &     5.2 $\pm$    1.8 \\
OF background      &    11 $\pm$    2  &     6.6 $\pm$    1.6  &     4.6 $\pm$    1.2 \\
$\W\Z$/$\Z\Z$ background   &    50 $\pm$   25  &     3.9 $\pm$    2.0  &     2.2 $\pm$    1.1 \\
\hline\hline
Total background   &  2359 $\pm$  737  &    43.4 $\pm$   11.4  &    12.0 $\pm$    2.4 \\
\hline\hline
Data         &  2416  &  47  &  7 \\
\hline\hline\\
Source  & $100 \le \MET < 150\GeV$ & $150 \le \MET < 200\GeV$ & $\MET \ge 200\GeV$ \\
\hline\hline
\zjets background &   1.7 $\pm$    0.6  &   0.4 $\pm$  0.2  &   0.20  $\pm$  0.09  \\
OF background      &   4.6 $\pm$    1.2  &   0.8 $\pm$  0.3  &   0.06 $\pm$  0.07  \\
$\W\Z$/$\Z\Z$ background   &   2.5 $\pm$    1.3  &   0.7 $\pm$  0.4  &   0.4  $\pm$  0.2 \\
\hline
Total background   &   8.8 $\pm$    1.8  &   1.9 $\pm$  0.5  &   0.7 $\pm$  0.3 \\
\hline\hline
Data         &  6  &  2  &  0  \\
\hline\hline
\end{tabular}

\end{center}
\end{table}

\section{Interpretations of the searches}
\label{conclusion}

In this section, we present the interpretation of our results.
Section~\ref{limit-trimet} presents the limits on the SMS of
Fig.~\ref{fig:charginos-slep} from the three-lepton search using the
$\MET$ shape (Section~\ref{trilepton-broad}).
Section~\ref{tri-ss-combine} presents the limits on the same SMS from
the three-lepton search using $\mdil$ and $\MT$
(Section~\ref{trilepton-targeted}), the same-sign dilepton search
(Section~\ref{dilepton}), and their combination.
Section~\ref{sec:bsm} presents the limits on the SMS of
Fig.~\ref{fig:charginos-wz} using results from Section~\ref{trilepton}
and from the $\W\Z + \MET$ analysis of Section~\ref{diboson}, as well
as limits on a GMSB model using results from the $\Z\Z + \MET$
analysis of Section~\ref{diboson} and the four-lepton results of
Ref.~\cite{SUS-11-013-paper}.  In all the search channels, the
observations agree with the expected background.

We present upper limits on the cross sections for pair production
of charginos and neutralinos.
All upper limits are computed at 95\%
confidence level (CL) using the \cls
criterion~\cite{Junk:1999kv,Read:2002hq} with choices in the
implementation following those in Ref.~\cite{ATLAS:1379837}.
Using the NLO cross section calculations from
Ref.~\cite{Beenakker:1999xh,Beenakker:1999xhErr,bib-NLO-NLL}, we also
evaluate 95\% CL exclusion curves.  The exclusion curves are shown
not only for their central values, but also when the NLO cross section
is varied by $\pm 1$ standard deviation ($\sigma$) of its
uncertainty \cite{bib-NLO-NLL}. In addition, we display the median expected
exclusion limit in an ensemble of experiments with background only,
as well as the uncertainty band that contains 68\% of the limits
in the ensembles.

\subsection{Limits on SMS from the search with three leptons using
\texorpdfstring{$\MET$}{MET} shape}
\label{limit-trimet}

\begin{figure} [!b]
\begin{center}
\subfigure[]{\includegraphics[width=0.49\textwidth]{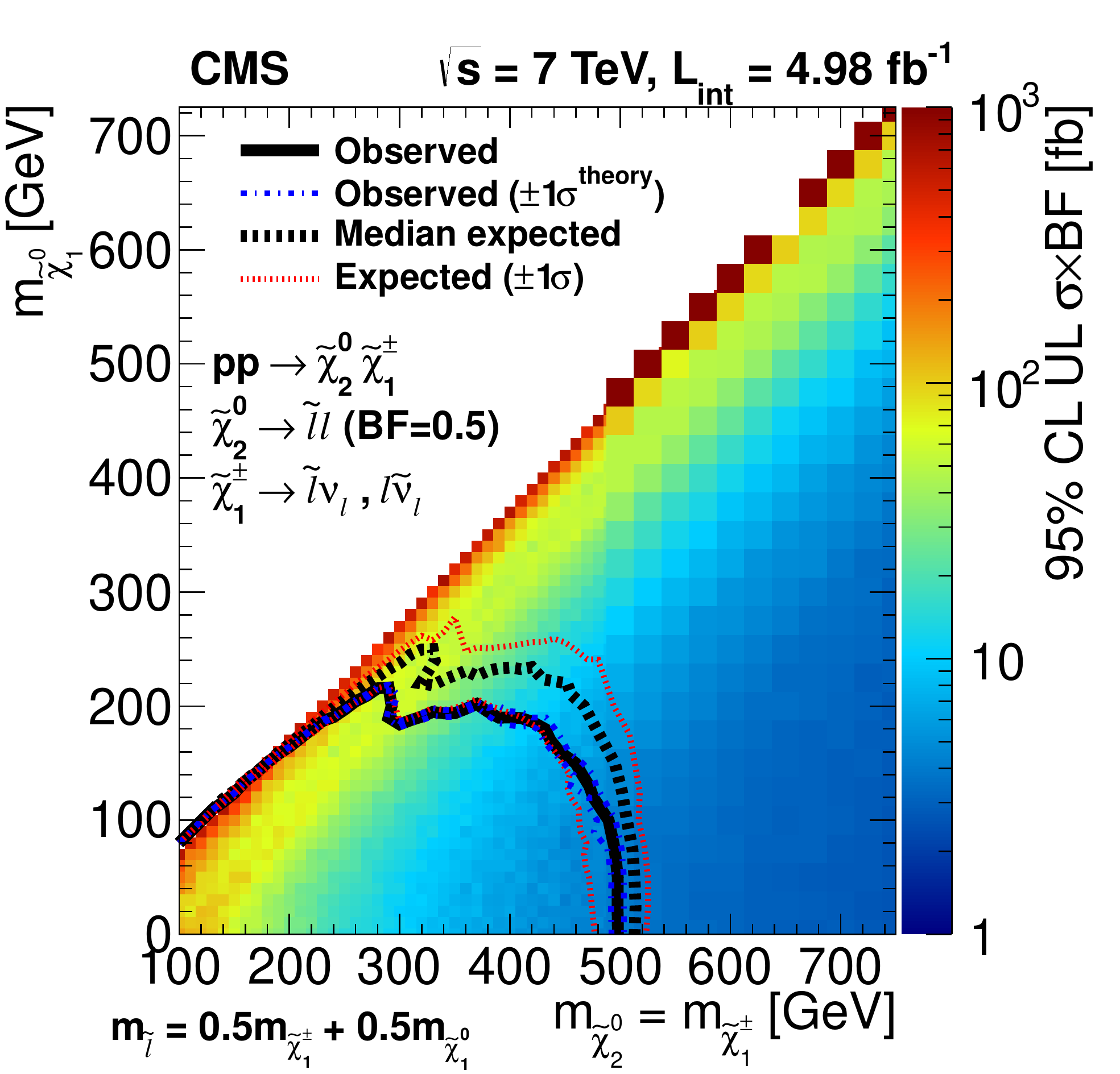} }
\subfigure[]{\includegraphics[width=0.49\textwidth]{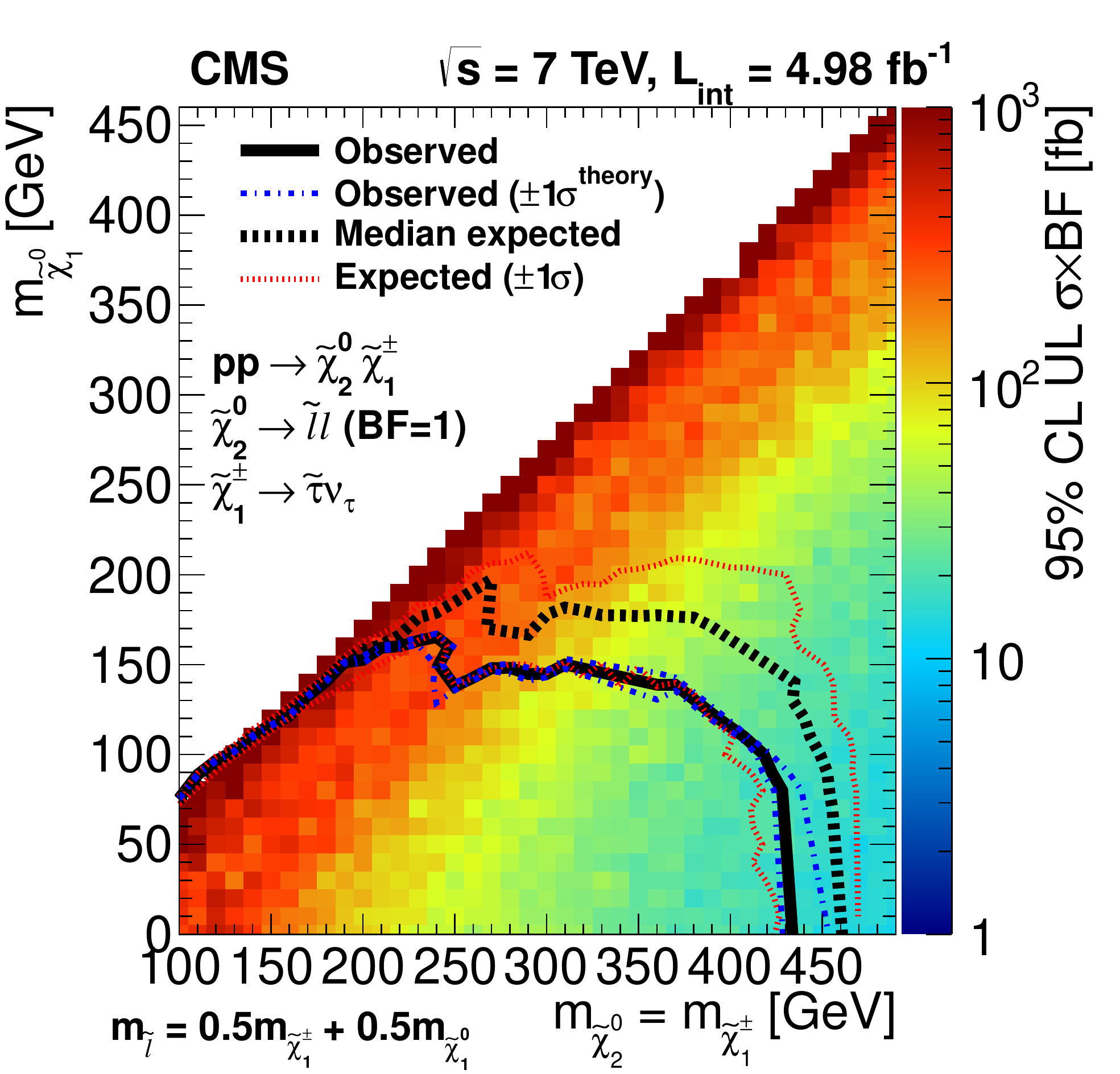} }
\caption{
\label{fig-sus-013-reinterp}
The shading in the $m_{\chiz_1}$ versus $m_{\chiz_2}$
($=m_{\chipm_1}$) plane indicates the 95\% CL upper limit on the
chargino-neutralino NLO production cross section times branching fraction
in (a) the flavor-democratic scenario, and (b) the $\tau$-enriched
scenario, based on the results of the three-lepton$+\MET$ search using
the data of Ref.~\citen{SUS-11-013-paper}.  The slepton mass is the
mean of the $\chiz_1$ and $\chipm_1$ masses, i.e., $\xslep=0.5$.  In
(a), the solid (dotted) contours bound the observed (expected) mass
region excluded at 95\% CL for a branching fraction of 50\%, as
appropriate for the three-lepton decay products in the
flavor-democratic scenario.  In (b), the same contours are for a
branching fraction of 100\%, as appropriate for the $\tau$-enriched
scenario, in which the final-state lepton from the chargino decay is
always the $\tau$ lepton.  }
\end{center}
\end{figure}

Figure~\ref{fig-sus-013-reinterp}(a) displays the 95\% CL upper limits
on the cross section times branching fraction in the $m_{\chiz_1}$
versus $m_{\chiz_2}$ ($=m_{\chipm_1}$) plane, with $\xslep=0.5$ in the
flavor-democratic scenario described in the Introduction.  The contour
bounds the excluded region in the plane assuming the NLO cross section
calculation and a 50\% branching fraction to three leptons, as
appropriate for this SMS.  Figure~\ref{fig-sus-013-reinterp}(b)
displays the corresponding limits for the $\tau$-enriched scenario.
The lower-sensitivity feature in the curve, noticeable where the
common mass $\mchi$ is approximately $100\GeV$ greater than
$m_{\chiz_1}$, corresponds to the phase space where the dilepton mass
has a high probability to be close to the $\Z$ mass, such that the
event is rejected.
\nopagebreak
\subsection{Limits on SMS from the search with three leptons, \texorpdfstring{$\mdil$, and $\MT$}{dilepton mass, and transverse mass},
and from same-sign dilepton searches}
\label{tri-ss-combine}

\begin{figure}[!p]
\begin{center}
\subfigure[]{\label{fig:UL_Model1}\includegraphics[scale=0.4]{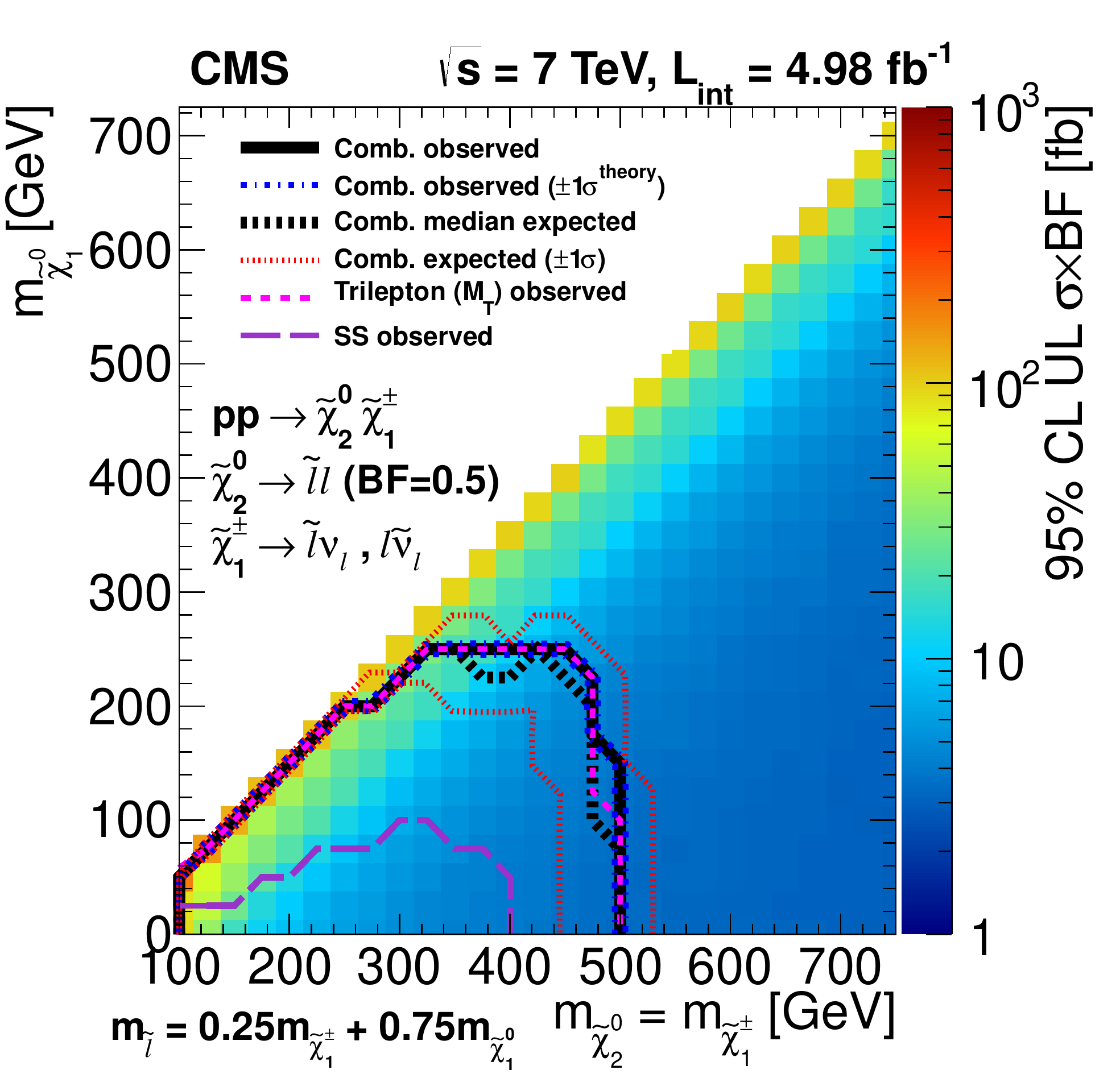} }
\subfigure[]{\label{fig:UL_Model2}\includegraphics[scale=0.4]{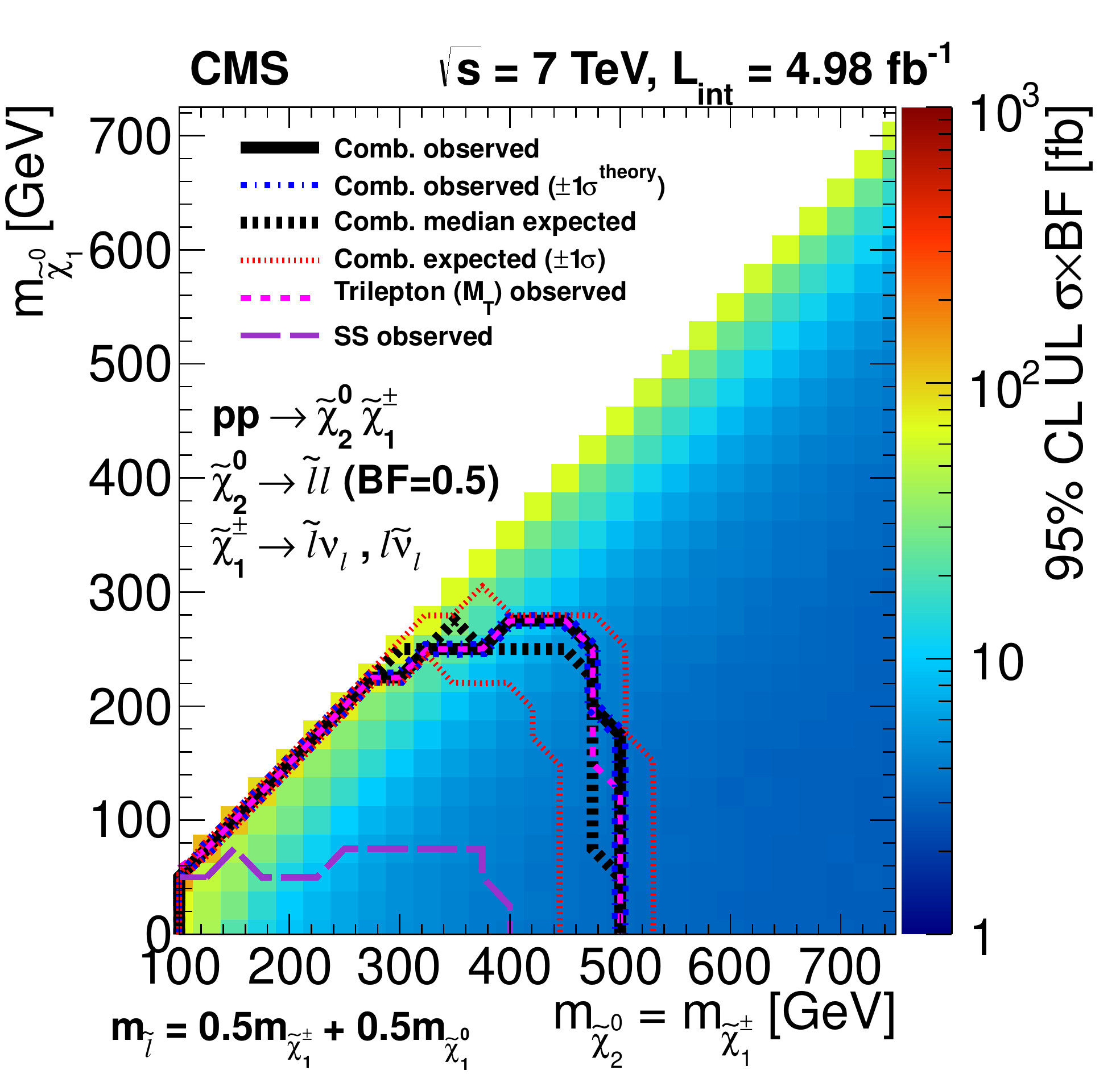} }\\
\subfigure[]{\label{fig:UL_Model3}\includegraphics[scale=0.4]{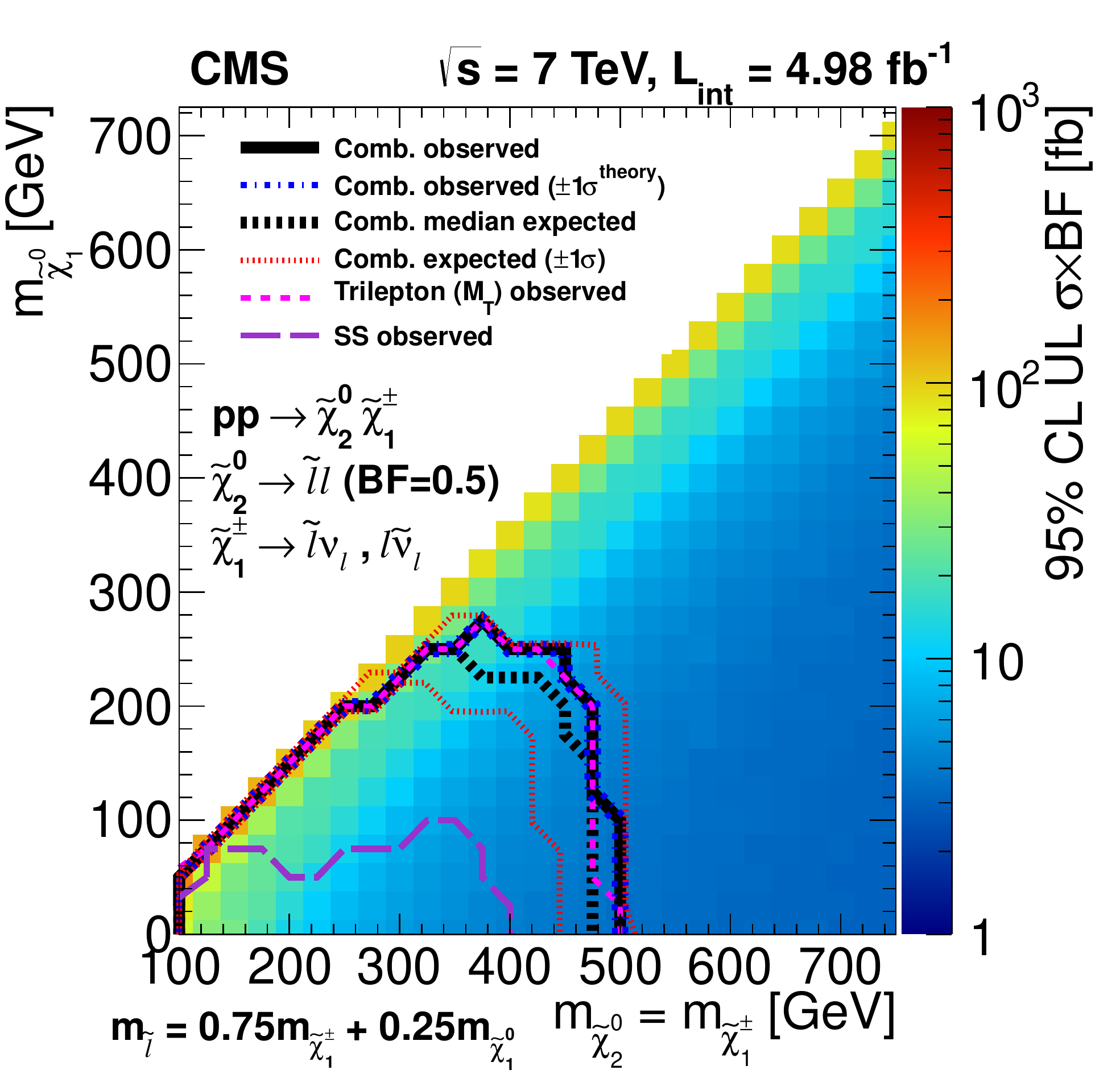} }
\caption{
The shading in the $m_{\chiz_1}$ versus $m_{\chiz_2}$
($=m_{\chipm_1}$) plane indicates the 95\% CL upper limit on the
chargino-neutralino production NLO cross section times branching fraction
in the flavor-democratic scenario, for the combined analysis of the
three-lepton search using $\mdil$ and $\MT$, and the same-sign
dilepton search.  The contours bound the mass regions excluded at 95\%
CL for a branching fraction of 50\%, as appropriate for the visible
decay products in this scenario. The contours based on the
observations are shown for the separate searches and for the
combination; in addition, the expected combined bound is shown.  The
three subfigures are the results for $\xslep$ set to (a) 0.25, (b) 0.50,
and (c) 0.75.
\label{fig:ULtriA}
}
\end{center}
\end{figure}

Figure~\ref{fig:ULtriA} displays, for three values of $\xslep$, the
95\% CL upper limit on the chargino-neutralino production cross
section times branching fraction in the flavor-democratic scenario,
derived from the results of the three-lepton search using $\MT$ and
$\mdil$ and those of the SS dilepton search.  The contours bound the
mass regions excluded at 95\% CL for a branching fraction of 50\%, as
appropriate for the visible decay products in this scenario. The
contours based on the observations are shown for the separate searches
and for the combination.  This search has slightly better sensitivity
than the complementary search based on the $\MET$ shape
(Fig.~\ref{fig-sus-013-reinterp}) in the region where the difference
between $\mchi$ and $m_{\chiz_1}$ is small, and slightly worse
sensitivity where this mass difference is large.

\begin{figure}[!p]
\begin{center}
\subfigure[]{\label{fig:UL_Model1B}\includegraphics[scale=0.4]{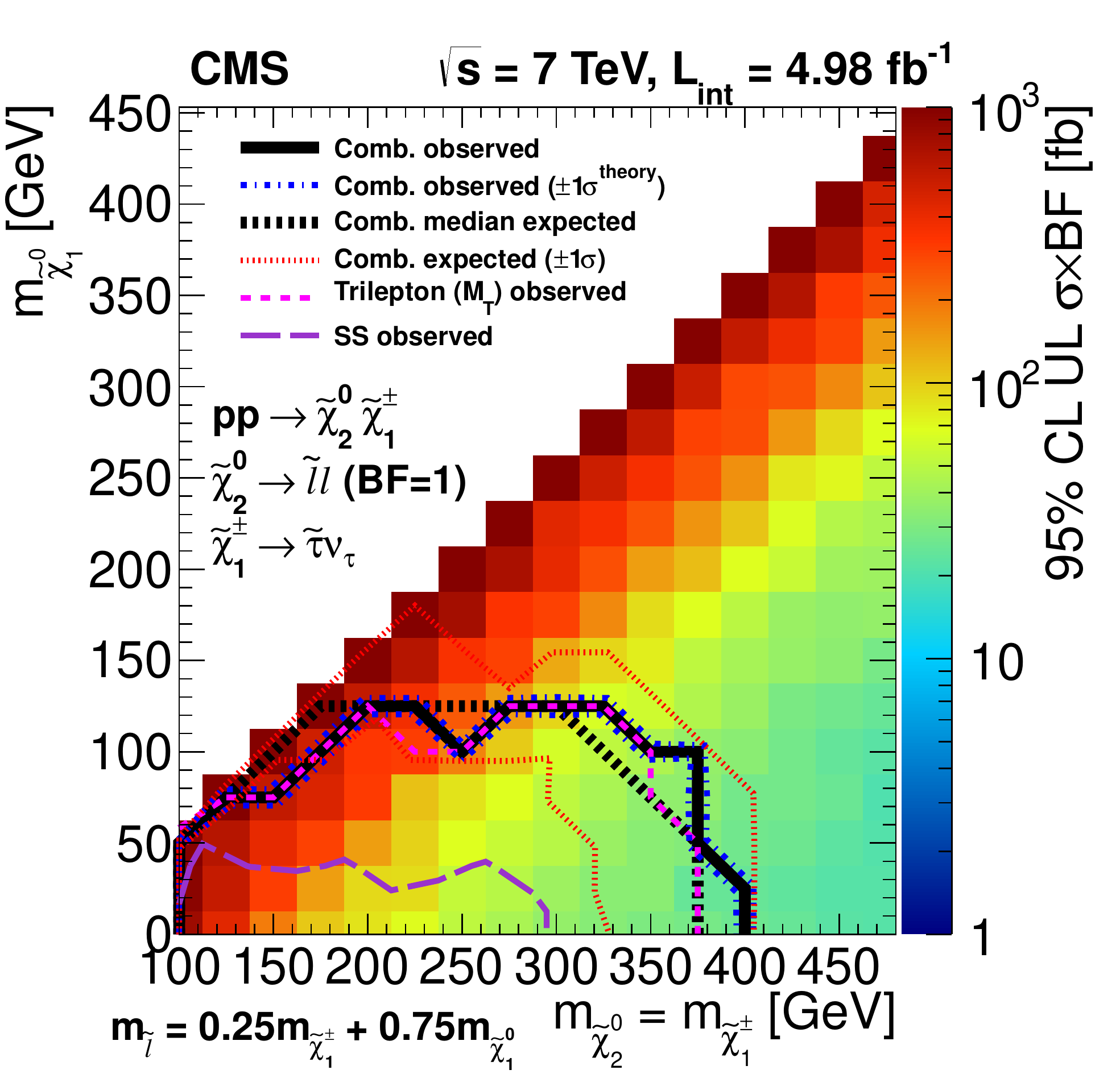} }
\subfigure[]{\label{fig:UL_Model2B}\includegraphics[scale=0.4]{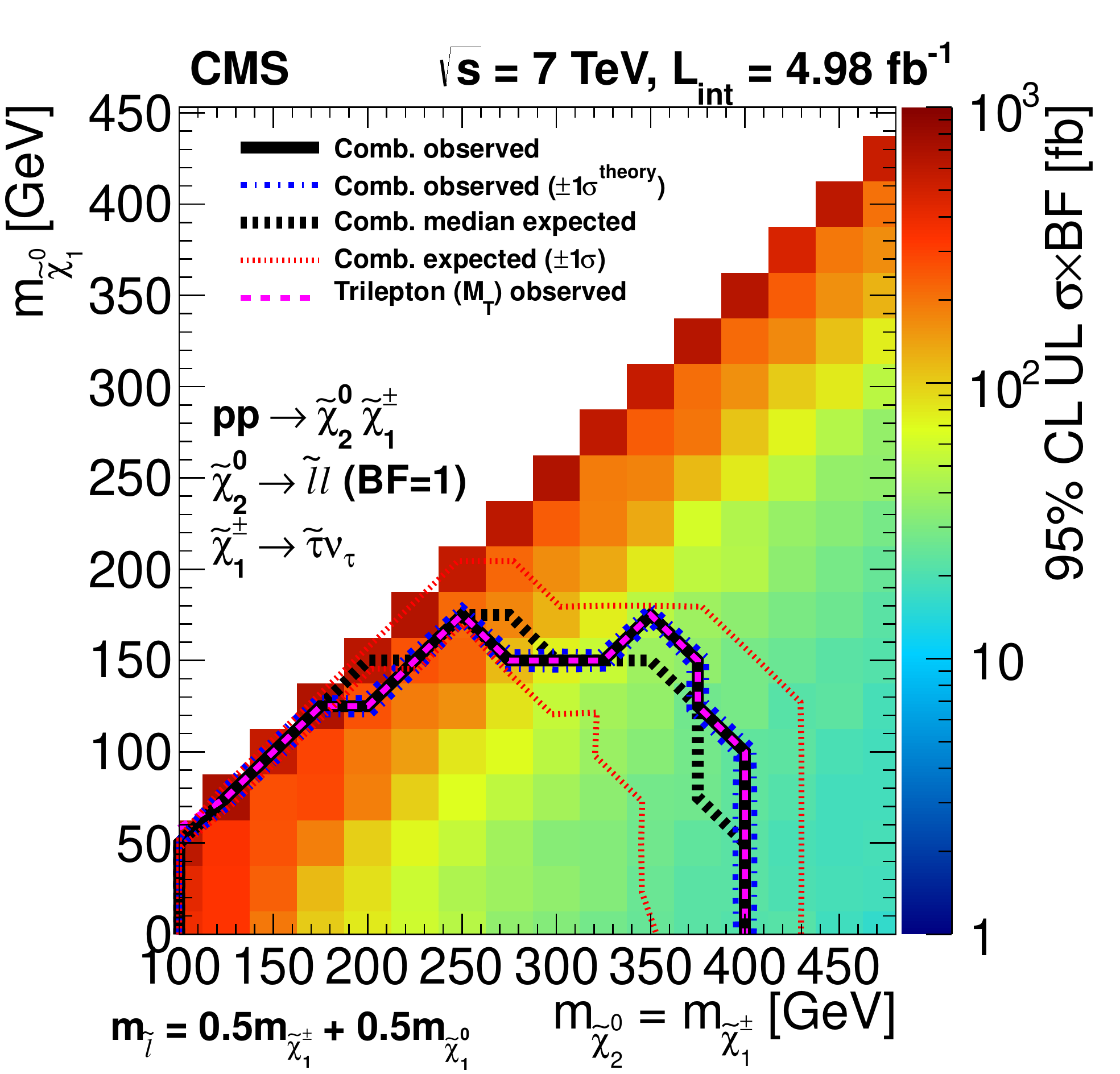} }\\
\subfigure[]{\label{fig:UL_Model3B}\includegraphics[scale=0.4]{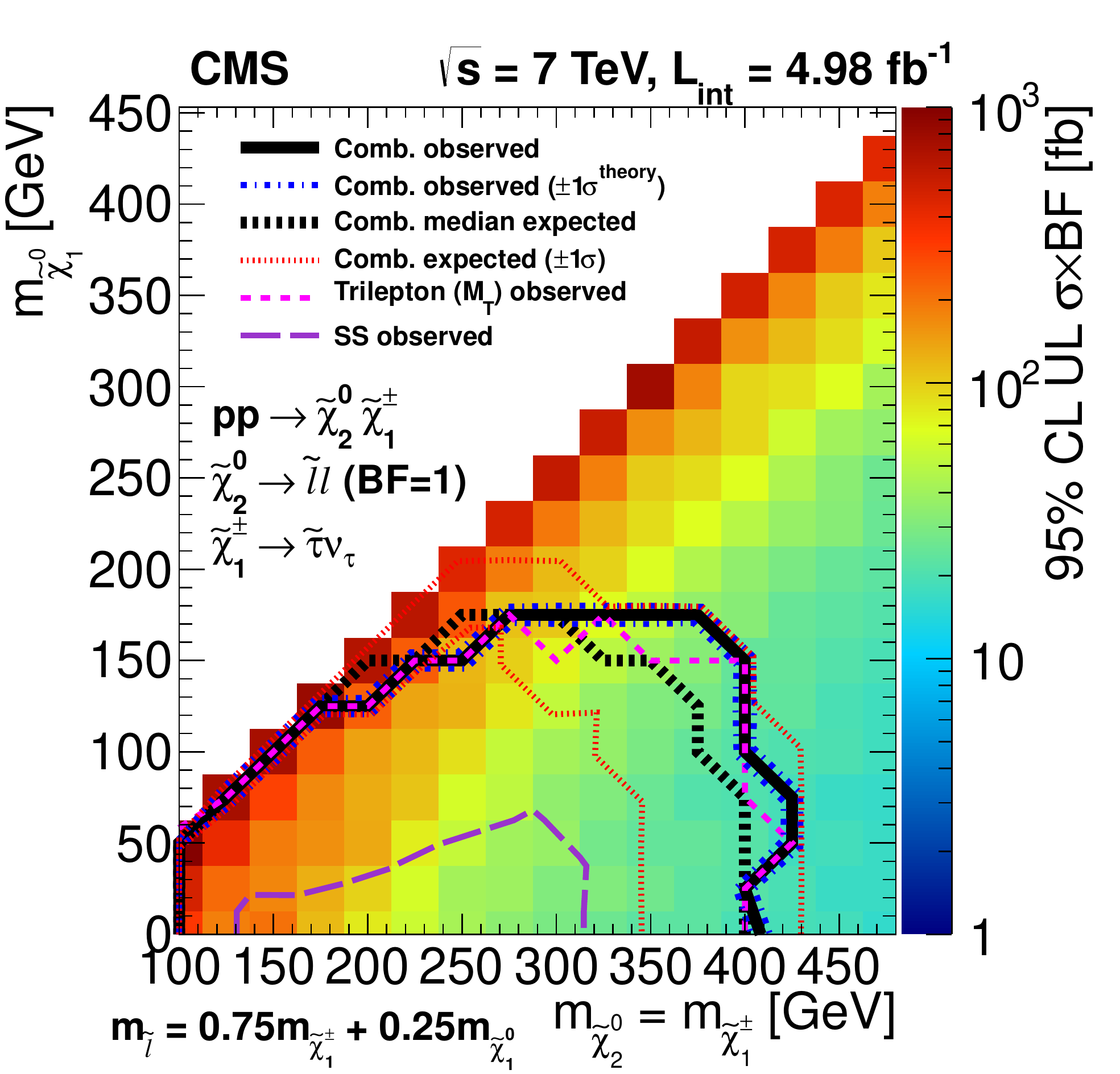} }
\caption{
For the $\tau$-enriched scenario, the results corresponding to those in
Fig.~\ref{fig:ULtriA}.
\label{fig:ULtriB}
}
\end{center}
\end{figure}

Figure~\ref{fig:ULtriB} presents the corresponding limits for the
$\tau$-enriched scenario.  As the SS dilepton search does not have
sensitivity for $\xslep=0.50$, there is no limit curve for this search
in Fig.~\ref{fig:ULtriB}(b).  In the other limit curves in both
Figs.~\ref{fig:ULtriA} and \ref{fig:ULtriB}, the increase in the
combined mass limit from incorporation of the SS dilepton search
ranges up to approximately $20\GeV$.

Appendix A provides a prescription for emulating the event selection
efficiency for this signature, in order to facilitate further
interpretation of the results in electroweak SUSY production scenarios
beyond the models considered in this paper.

\subsection{Limits on SMS and GMSB from the
\texorpdfstring{$\W\Z/\Z\Z + \MET$}{WZ/ZZ + MET} final state with two or
more leptons}
\label{sec:bsm}
We calculate upper limits on the cross sections for pair production of
charginos and neutralinos times branching fractions into the \wzmet
and \zzmet final states as a function of the chargino and neutralino
masses.  In calculating these limits, the uncertainties related to jet
and \MET quantities (jet multiplicity, dijet mass, and \MET) vary
significantly across the model space, and are addressed separately at
each point, taking into account the bin-to-bin migration of signal
events.  The limits in Section~\ref{app:combo_trilepton} are presented
in the context of the SMS of Fig.~\ref{fig:charginos-wz}(a) with 100\%
branching fractions of the chargino (neutralino) to $\W+\chiz_1$
($\Z+\chiz_1$).  The wino-like cross section with coupling
$g\gamma^{\mu}$ is assumed.  As the present data do not have
sufficient sensitivity to explore the SMS of
Fig.~\ref{fig:charginos-wz}(b), the limits in Section~\ref{app:combo}
are presented in the context of a gauge-mediated symmetry breaking
(GMSB) \Z-enriched higgsino
model~\cite{Matchev:1999ft,Meade:2009qv,ref:ewkino} that has a large
branching fraction to the \zzmet final state. In this scenario, the
LSP is a very light gravitino (mass $\leq$ 1 keV).

\subsubsection{Limits on SMS with on-shell \texorpdfstring{$\W$}{W} and
\texorpdfstring{$\Z$}{Z} from \texorpdfstring{$\W\Z + \MET$}{WZ + MET }
and three-lepton analyses}
\label{app:combo_trilepton}

For limits on the SMS of Fig.~\ref{fig:charginos-wz}(a) with on-shell
$\W$ and $\Z$ bosons, we combine the results of the \wzzmet analysis
and the three-lepton analysis of Section~\ref{trilepton-targeted}.
From the \wzzmet analysis, we use the results in exclusive
\MET regions, as summarized in Table~\ref{resulttableex}.  For the
three-lepton analysis, we use the results in Table~\ref{tab:limits}.
The three-lepton region with the broadest sensitivity is Region III,
the on-\Z, high-$\MT$ region. If the difference between the common
mass $\mchi$ and $m_{\chiz_1}$ is small, then a significant fraction
of the signal events fall below the \Z mass window so that other
signal regions contribute as well, in particular Region I (below-\Z,
low-$\MT$ region).  Region VI is not used directly in the fit, in
order to facilitate the combination and to avoid using this region to
constrain the $\W\Z$ yield in the \wzzmet\ analysis, where the
kinematic selection is very different since it includes jet
requirements.  Instead, a scaling factor of $1.1\pm0.1$ is applied to
the $\W\Z$ yield in Regions I-V, based on the data/simulation
comparison in Region VI.

In the combination, the common signal-related systematic uncertainties
for luminosity, jet energy scale, lepton identification, trigger efficiency, and
misidentification of light-flavor jets as \bjetsnohyphen are
considered to be 100\%
correlated.  For backgrounds, the only common systematic uncertainty
is that for the $\W\Z/\Z\Z$ simulation, which is treated as 100\% correlated.  No
events in the data pass both signal selections.  For the backgrounds,
the overlap in the control sample is less than 1\%.  Thus the two
selections are treated as independent.

\begin{figure}[t!]
  \begin{center}
    \includegraphics[scale=0.4]{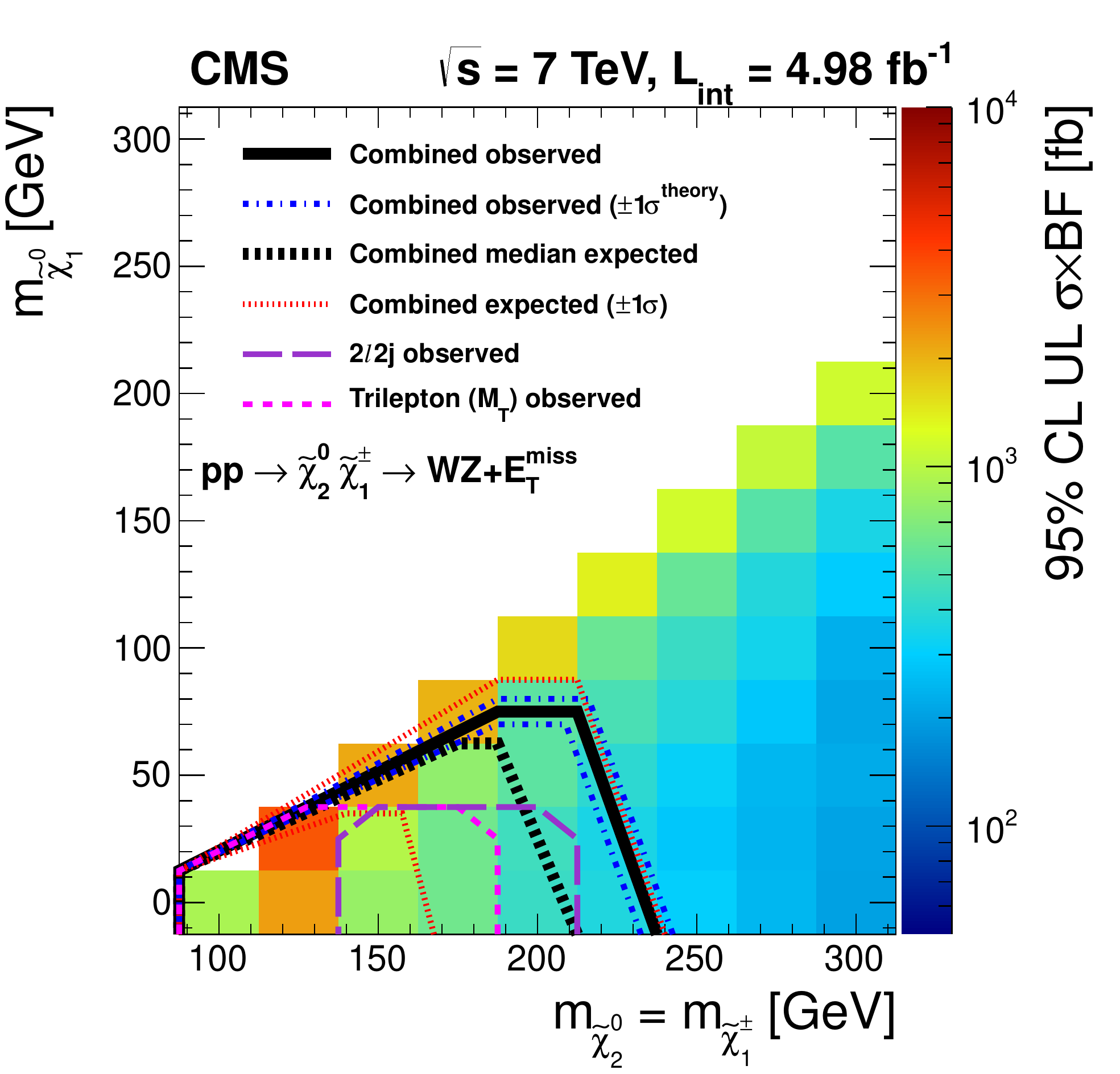}
    \caption{
      \label{fig:combo_trilepton}
Interpretation of the \wzmet and three-lepton results in the context
of the $\W\Z$ SMS.  The \wzmet observed, three-lepton observed,
combined observed, and combined expected contours are indicated.
    }
  \end{center}
\end{figure}

Figure~\ref{fig:combo_trilepton} displays the observed limits for the
two individual analyses and the combination.  For large $\mchi$, the
\wzzmet analysis has higher sensitivity due to the large hadronic
branching fractions of the $\W$ and $\Z$ bosons. At lower $\mchi$, the
signal events do not have large $\MET$, resulting in a loss of signal
region acceptance for the \wzzmet analysis. In this region, the
background suppression provided by the requirement of a third lepton
leads to better sensitivity for the three-lepton analysis.

\subsubsection{Limits on a \texorpdfstring{$\Z$}{Z}-enriched GMSB model from
\texorpdfstring{$\Z\Z + \MET$}{ZZ + MET }
and four-lepton search}

\label{app:combo}

For the SMS of Fig.~\ref{fig:charginos-wz}(b) with two on-shell $\Z$
bosons, the present data do not exclude any region of $\mchi$, and are
therefore not sensitive to a scenario in which neutralino pair
production is the sole production mechanism.  However, the \zzmet
signature can be enhanced in scenarios in which additional mechanisms,
such as chargino-chargino and chargino-neutralino production, also
contribute. This is the case in a GMSB \Z-enriched higgsino
model~\cite{Matchev:1999ft,Meade:2009qv,ref:ewkino}.

In this scenario, the LSP is a nearly massless gravitino, the
next-to-lightest SUSY particle is a \Z-enriched higgsino $\chiz_1$,
and the $\chipm_1$ is nearly mass degenerate with the $\chiz_{1}$.  We
set the gaugino mass parameters $M_1$ and $M_2$ to $M_1=M_2=1\TeV$,
the ratio of Higgs expectation values $\tan\beta$ to $\tan\beta=2$,
and then explore variable Higgsino mass parameters. The masses of the
$\chiz_1$ and $\chipm_1$ are controlled by the parameter $\mu$, with
$m_{\chiz_1} \approx m_{\chipm_1} \approx \mu$.  Hence the $\chipm_1$
decays to $\chiz_{1}$ and to low-\PT SM particles that escape
detection. Thus, all production mechanisms (chargino-chargino,
chargino-neutralino, and neutralino-neutralino) lead to a pair of
$\chiz_{1}$ particles in the final state, and the branching fraction
to the \zzmet final state
is large (varying from 100\% at $\mu=130$~\GeV to 85\% at
$\mu=410$~\GeV).  Mainly because of the mix of production mechanisms,
the kinematic distributions of this model are slightly different than
those expected in a pure neutralino-pair production scenario, leading
to different signal acceptances.

We combine the results of the \wzzmet analysis of Section~\ref{diboson}
with independent results for the four-lepton channel analysis of
Ref.~\citen{SUS-11-013-paper} to further restrict the GMSB scenario.
The two selections have negligible overlap, and are thus treated as
independent in the combination.

Table~\ref{multilepresults} summarizes the relevant results from
Ref.~\citen{SUS-11-013-paper}, with the high-\HT and low-\HT regions
of that study combined.  All samples contain four leptons, including
an OSSF lepton pair whose mass is consistent with the $\Z$ boson mass,
with separate entries for events with \MET above or below $50\GeV$,
and for events with zero or one hadronically decaying $\tau$ lepton
candidate ($\tau_h$).

\begin{table}[t!]
\begin{center}
\topcaption{\label{multilepresults}
Summary of the results from the multilepton analysis of
Ref.~\citen{SUS-11-013-paper} used as input to the combined limit on
the GMSB model. All categories have four leptons including an OSSF
pair consistent with a $\Z$ boson; $N(\tau_h)$ denotes the number of
these leptons that are identified as hadronically decaying $\tau$
leptons.  Uncertainties include statistical and systematic
contributions.  }
\begin{tabular}{lcc}
\hline\hline
Signal Region    &   Expected Background & Observed Yield \\
\hline
$N(\tau_h)=0$  , \MET $\ge$ 50 GeV & $1.0 \pm 0.2$ &  1 \\
$N(\tau_h)=0$  , \MET $<$ 50 GeV & $38  \pm  15$ & 34 \\
$N(\tau_h)=1$  , \MET $\ge$ 50 GeV & $2.6 \pm 0.7$ &  4 \\
$N(\tau_h)=1$  , \MET $<$ 50 GeV & $18.0  \pm 5.2$ & 20 \\
\hline\hline
\end{tabular}
\end{center}
\end{table}

\begin{figure}[htb]
  \begin{center}
    \includegraphics[width=0.5\textwidth]{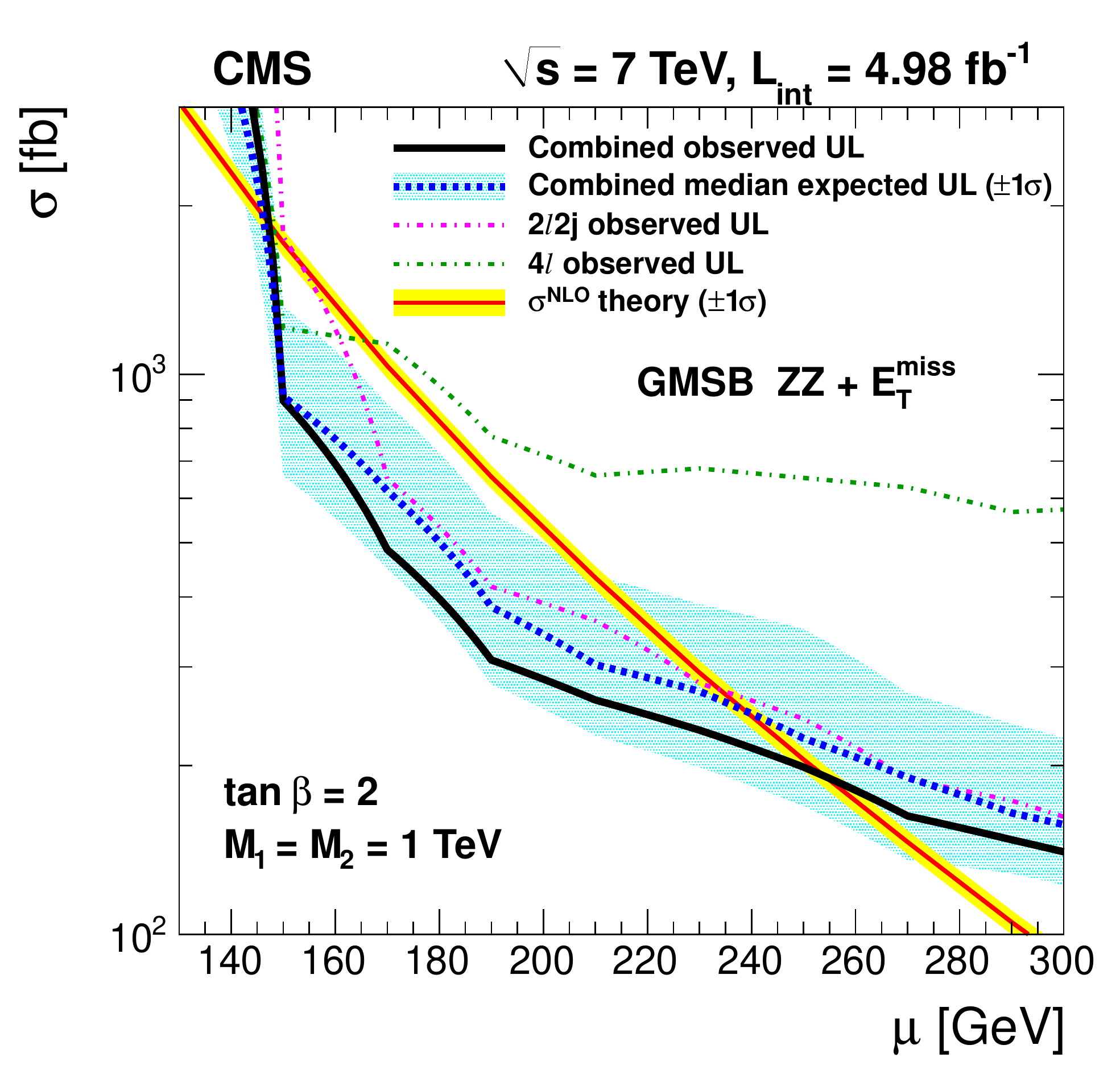}
\caption{
\label{fig:combo}
Interpretation of the results for the \zzmet (with two leptons and two
jets) analysis and the results of the four-lepton analysis
from Ref.~\cite{SUS-11-013-paper} in the
context of the GMSB model described in the text.  The NLO cross section
upper limits are presented for the \zzmet observed, multilepton
observed, the combined observed, and the combined expected
results. The theory prediction for the cross section is also
presented.  The median expected limits, their $\pm1\sigma$ variations,
and the $\pm1\sigma$ band on the theory curve are as described at the
beginning of Section~\ref{conclusion}.
}
\end{center}
\end{figure}

The results of the individual and combined exclusions are displayed in
Fig.~\ref{fig:combo}.  As in Section~\ref{app:combo_trilepton}, the
\wzzmet and the multilepton analysis are complementary,
with the four-lepton analysis having greater (less) sensitivity than
the \wzzmet analysis at small (large) values of $\mu$.  By combining
the two analyses, we exclude the range of $\mu$ between 148 and 248
\GeV.

\subsection{Summary of excluded masses for chargino-neutralino pair production}
\label{limitsummary}

\begin{figure}[t!]
  \begin{center}
    \includegraphics[width=0.6\textwidth]{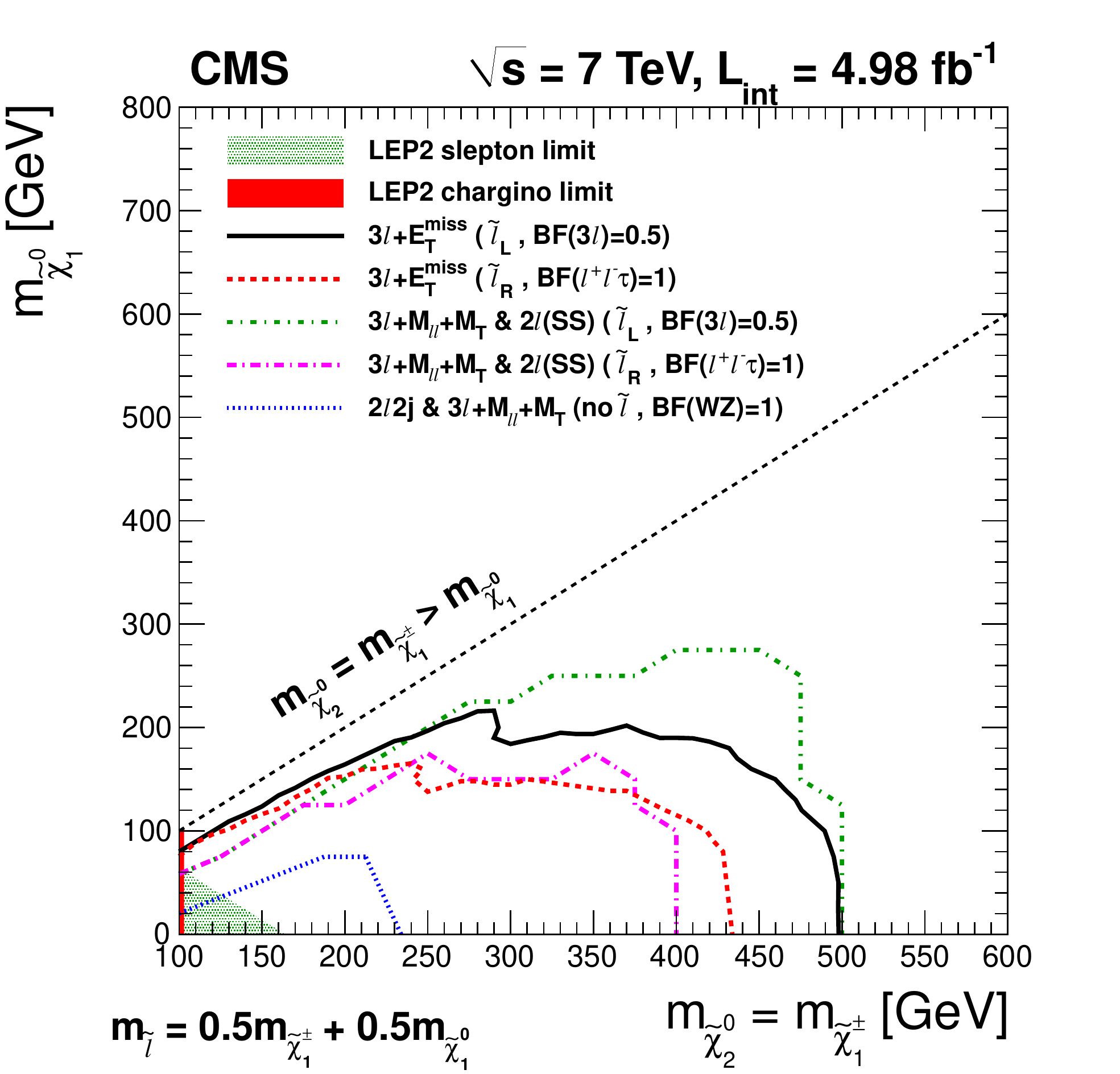}
\caption{ \label{fig:summary} Summary of the excluded regions in
the $m_{\chiz_1}$ versus $m_{\chiz_2}$ ($=m_{\chipm_1}$) plane for:
the three-lepton$+\MET$ search (Sections~\ref{trilepton-broad} and
\ref{limit-trimet}), separately for the ${\slep}_L$ and ${\slep}_R$
scenarios;
the combination (Section~\ref{tri-ss-combine}) of the three-lepton
analysis based on $\mdil$ and $\MT$
(Section~\ref{trilepton-targeted}) with the SS dilepton analysis
(Section~\ref{dilepton}), separately for the ${\slep}_L$ and
${\slep}_R$ scenarios;
and the combination (Section~\ref{app:combo_trilepton}) of the diboson
analysis with two leptons and two jets (Section~\ref{diboson}) with
the three-lepton analysis based on $\mdil$ and $\MT$
(Section~\ref{trilepton-targeted}), for the \wzmet model.
Regions excluded by searches at LEP2 for sleptons and charginos are
also indicated.  The implied branching fractions introduced in
Section~\ref{introduction} are noted in the legend. For models with
intermediate sleptons (including the LEP2 slepton limit), the
interpretations correspond to $\xslep=0.5$.
} \end{center}
\end{figure}

Figure~\ref{fig:summary} displays a summary of the excluded regions in
the chargino-neutralino production scenarios considered above. Also
displayed are the exclusion curves at 95\% CL from searches at
LEP2~\cite{delphi,Heister:2002jca,PDG}, which excluded $m_{\slep}<82
\GeV$ and $m_{\chipm_1}< 103 \GeV$.  The results in this paper probe
the production of charginos and neutralinos with masses up to
approximately 200 to $500\GeV$, depending on the decay modes of these
particles.

\section{Summary}

This paper presents searches for supersymmetric charginos and
neutralinos.  While a number of previous studies at the LHC have
focused on strongly coupled supersymmetric particles, this paper is
one of the first to focus on the electroweak sector of supersymmetry.
The searches performed here explore final states with exactly
three leptons using transverse mass and lepton-pair invariant mass,
two same-sign leptons, and two opposite-sign leptons and two jets.
The results of a published search for new physics in the final state
of three or more leptons are reinterpreted in the context of
electroweak supersymmetry.  No excesses above the standard model
expectations are observed.  The results are used to exclude a range of
chargino and neutralino masses from approximately 200 to $500 \GeV$
in the context of models that assume large branching fractions of
charginos and neutralinos to leptons and vector bosons.

\clearpage

\section*{Acknowledgements}

We thank David Shih for useful discussions and help with
implementation of the $\Z$-enriched GMSB model of
Section~\ref{app:combo}.

\hyphenation{Bundes-ministerium Forschungs-gemeinschaft Forschungs-zentren} We congratulate our colleagues in the CERN accelerator departments for the excellent performance of the LHC machine. We thank the technical and administrative staff at CERN and other CMS institutes. This work was supported by the Austrian Federal Ministry of Science and Research; the Belgian Fonds de la Recherche Scientifique, and Fonds voor Wetenschappelijk Onderzoek; the Brazilian Funding Agencies (CNPq, CAPES, FAPERJ, and FAPESP); the Bulgarian Ministry of Education and Science; CERN; the Chinese Academy of Sciences, Ministry of Science and Technology, and National Natural Science Foundation of China; the Colombian Funding Agency (COLCIENCIAS); the Croatian Ministry of Science, Education and Sport; the Research Promotion Foundation, Cyprus; the Ministry of Education and Research, Recurrent financing contract SF0690030s09 and European Regional Development Fund, Estonia; the Academy of Finland, Finnish Ministry of Education and Culture, and Helsinki Institute of Physics; the Institut National de Physique Nucl\'eaire et de Physique des Particules~/~CNRS, and Commissariat \`a l'\'Energie Atomique et aux \'Energies Alternatives~/~CEA, France; the Bundesministerium f\"ur Bildung und Forschung, Deutsche Forschungsgemeinschaft, and Helmholtz-Gemeinschaft Deutscher Forschungszentren, Germany; the General Secretariat for Research and Technology, Greece; the National Scientific Research Foundation, and National Office for Research and Technology, Hungary; the Department of Atomic Energy and the Department of Science and Technology, India; the Institute for Studies in Theoretical Physics and Mathematics, Iran; the Science Foundation, Ireland; the Istituto Nazionale di Fisica Nucleare, Italy; the Korean Ministry of Education, Science and Technology and the World Class University program of NRF, Korea; the Lithuanian Academy of Sciences; the Mexican Funding Agencies (CINVESTAV, CONACYT, SEP, and UASLP-FAI); the Ministry of Science and Innovation, New Zealand; the Pakistan Atomic Energy Commission; the Ministry of Science and Higher Education and the National Science Centre, Poland; the Funda\c{c}\~ao para a Ci\^encia e a Tecnologia, Portugal; JINR (Armenia, Belarus, Georgia, Ukraine, Uzbekistan); the Ministry of Education and Science of the Russian Federation, the Federal Agency of Atomic Energy of the Russian Federation, Russian Academy of Sciences, and the Russian Foundation for Basic Research; the Ministry of Science and Technological Development of Serbia; the Secretar\'{\i}a de Estado de Investigaci\'on, Desarrollo e Innovaci\'on and Programa Consolider-Ingenio 2010, Spain; the Swiss Funding Agencies (ETH Board, ETH Zurich, PSI, SNF, UniZH, Canton Zurich, and SER); the National Science Council, Taipei; the Thailand Center of Excellence in Physics, the Institute for the Promotion of Teaching Science and Technology and National Electronics and Computer Technology Center; the Scientific and Technical Research Council of Turkey, and Turkish Atomic Energy Authority; the Science and Technology Facilities Council, UK; the US Department of Energy, and the US National Science Foundation.

Individuals have received support from the Marie-Curie programme and the European Research Council (European Union); the Leventis Foundation; the A. P. Sloan Foundation; the Alexander von Humboldt Foundation; the Austrian Science Fund (FWF); the Belgian Federal Science Policy Office; the Fonds pour la Formation \`a la Recherche dans l'Industrie et dans l'Agriculture (FRIA-Belgium); the Agentschap voor Innovatie door Wetenschap en Technologie (IWT-Belgium); the Ministry of Education, Youth and Sports (MEYS) of Czech Republic; the Council of Science and Industrial Research, India; the Compagnia di San Paolo (Torino); and the HOMING PLUS programme of Foundation for Polish Science, cofinanced from European Union, Regional Development Fund.

\bibliography{auto_generated}   % will be created by the tdr script.

\providecommand{\href}[2]{#2}\begingroup\raggedright\begin{thebibliography}{10}%
\makeatletter
\providecommand{\hrefCMSnoop }[0]{\@secondoftwo}%
\makeatother
\providecommand{\doi}{\texttt{doi:}\begingroup \urlstyle{tt}\Url}

\bibitem{Golfand:1971iw}
\href {http://www.jetpletters.ac.ru/ps/1584/article_24309.pdf} {Y.~A. Golfand
  and E.~P. Likhtman, ``Extension of the algebra of Poincare group generators
  and violation of P invariance'',} \textit{ JETP Lett.} \textbf{ 13} (1971)
323.
%%CITATION = JTPLA,13,323;%%.

\bibitem{Ramond:1971gb}
\hrefCMSnoop {} {P.~Ramond, ``Dual theory for free fermions'',} \textit{ Phys.
  Rev. D} \textbf{ 3} (1971) 2415,
\href{http://dx.doi.org/10.1103/PhysRevD.3.2415}{\doi{10.1103/PhysRevD.3.2415}}.
%%CITATION = PHRVA,D3,2415;%%.

\bibitem{Neveu:1971rx}
\hrefCMSnoop {} {A.~Neveu and J.~H. Schwarz, ``{Factorizable dual model of
  pions}'',} \textit{ Nucl. Phys. B} \textbf{ 31} (1971) 86,
\href{http://dx.doi.org/10.1016/0550-3213(71)90448-2}{\doi{10.1016/0550-3213(71)90448-2}}.
%%CITATION = NUPHA,B31,86;%%.

\bibitem{Neveu:1971iv}
\hrefCMSnoop {} {A.~Neveu and J.~H. Schwarz, ``Quark model of dual pions'',}
  \textit{ Phys. Rev. D} \textbf{ 4} (1971) 1109,
\href{http://dx.doi.org/10.1103/PhysRevD.4.1109}{\doi{10.1103/PhysRevD.4.1109}}.
%%CITATION = PHRVA,D4,1109;%%.

\bibitem{Volkov:1973ix}
\hrefCMSnoop {} {D.~V. Volkov and V.~P. Akulov, ``Is the neutrino a goldstone
  particle?'',} \textit{ Phys. Lett. B} \textbf{ 46} (1973) 109,
\href{http://dx.doi.org/10.1016/0370-2693(73)90490-5}{\doi{10.1016/0370-2693(73)90490-5}}.
%%CITATION = PHLTA,B46,109;%%.

\bibitem{Wess:1973kz}
\hrefCMSnoop {} {J.~Wess and B.~Zumino, ``A lagrangian model invariant under
  supergauge transformations'',} \textit{ Phys. Lett. B} \textbf{ 49} (1974)
  52,
\href{http://dx.doi.org/10.1016/0370-2693(74)90578-4}{\doi{10.1016/0370-2693(74)90578-4}}.
%%CITATION = PHLTA,B49,52;%%.

\bibitem{Wess:1974tw}
\hrefCMSnoop {} {J.~Wess and B.~Zumino, ``Supergauge transformations in
  four-dimensions'',} \textit{ Nucl. Phys. B} \textbf{ 70} (1974) 39,
\href{http://dx.doi.org/10.1016/0550-3213(74)90355-1}{\doi{10.1016/0550-3213(74)90355-1}}.
%%CITATION = NUPHA,B70,39;%%.

\bibitem{Dicus:1983cb}
\hrefCMSnoop {} {D.~A. Dicus, S.~Nandi, and X.~Tata, ``{W} decay in
  supergravity {GUTS}'',} \textit{ Phys. Lett. B} \textbf{ 129} (1983) 451,
\href{http://dx.doi.org/10.1016/0370-2693(83)90138-7}{\doi{10.1016/0370-2693(83)90138-7}}.
%%CITATION = PHLTA,B129,451;%%.

\bibitem{Chamseddine:1983eg}
\hrefCMSnoop {} {A.~H. Chamseddine, P.~Nath, and R.~L. Arnowitt, ``Experimental
  signals for supersymmetric decays of the {W} and {Z} bosons'',} \textit{
  Phys. Lett. B} \textbf{ 129} (1983) 445,
\href{http://dx.doi.org/10.1016/0370-2693(83)90137-5}{\doi{10.1016/0370-2693(83)90137-5}}.
%%CITATION = PHLTA,B129,445;%%.

\bibitem{Baer:1985at}
\hrefCMSnoop {} {H.~Baer and X.~Tata, ``Multi-lepton signals from {W}$^{\pm}$
  and {Z}$^0$ decays to gauginos at $\bar{\mathrm{p}}\mathrm{p}$ colliders'',}
  \textit{ Phys. Lett. B} \textbf{ 155} (1985) 278,
\href{http://dx.doi.org/10.1016/0370-2693(85)90654-9}{\doi{10.1016/0370-2693(85)90654-9}}.
%%CITATION = PHLTA,B155,278;%%.

\bibitem{Nath:1987sw}
\hrefCMSnoop {} {P.~Nath and R.~L. Arnowitt, ``{Supersymmetric signals at the
  Tevatron}'',} \textit{ Mod. Phys. Lett. A} \textbf{ 2} (1987) 331,
\href{http://dx.doi.org/10.1142/S0217732387000446}{\doi{10.1142/S0217732387000446}}.
%%CITATION = MPLAE,A2,331;%%.

\bibitem{Baer:1994nr}
H.~Baer\hrefCMSnoop {} { {et~al.}, ``{Trileptons from chargino - neutralino
  production at the CERN Large Hadron Collider}'',} \textit{ Phys. Rev. D}
  \textbf{ 50} (1994) 4508,
  \href{http://dx.doi.org/10.1103/PhysRevD.50.4508}{\doi{10.1103/PhysRevD.50.4508}}.

\bibitem{Baer:1995va}
H.~Baer\hrefCMSnoop {} { {et~al.}, ``{Signals for minimal supergravity at the
  CERN Large Hadron Collider. II: multilepton channels}'',} \textit{ Phys. Rev.
  D} \textbf{ 53} (1996) 6241,
  \href{http://dx.doi.org/10.1103/PhysRevD.53.6241}{\doi{10.1103/PhysRevD.53.6241}}.

\bibitem{CMS-PAS-SMP-12-008}
\href {http://cdsweb.cern.ch/record/1434360} {{ CMS} Collaboration, ``Absolute
  calibration of the luminosity measurement at {CMS}: winter 2012 update'',}
  CMS Physics Analysis Summary CMS-PAS-SMP-12-008, (2012).

\bibitem{aleph}
\hrefCMSnoop {} {{ ALEPH} Collaboration, ``Absolute mass lower limit for the
  lightest neutralino of the MSSM from e$^+$e$^-$ data at $\sqrt{s}$ up to 209
  GeV'',} \textit{ Phys. Lett. B} \textbf{ 583} (2004) 247,
  \href{http://dx.doi.org/10.1016/j.physletb.2003.12.066}{\doi{10.1016/j.physletb.2003.12.066}}.
  See also references therein.

\bibitem{Heister:2002jca}
\hrefCMSnoop {} {{ ALEPH} Collaboration, ``Absolute lower limits on the masses
  of selectrons and sneutrinos in the {MSSM}'',} \textit{ Phys. Lett. B}
  \textbf{ 544} (2002) 73,
  \href{http://dx.doi.org/10.1016/S0370-2693(02)02471-1}{\doi{10.1016/S0370-2693(02)02471-1}}.

\bibitem{delphi}
\hrefCMSnoop {} {{ DELPHI} Collaboration, ``Searches for supersymmetric
  particles in e$^+$e$^-$ collisions up to 208 GeV and interpretation of the
  results within the MSSM'',} \textit{ Eur. Phys. J. C} \textbf{ 31} (2003)
  421,
  \href{http://dx.doi.org/10.1140/epjc/s2003-01355-5}{\doi{10.1140/epjc/s2003-01355-5}}.
  See also references therein.

\bibitem{l3}
\hrefCMSnoop {} {{ L3} Collaboration, ``Search for scalar leptons and scalar
  quarks at LEP'',} \textit{ Phys. Lett. B} \textbf{ 580} (2004) 37,
  \href{http://dx.doi.org/10.1016/j.physletb.2003.10.010}{\doi{10.1016/j.physletb.2003.10.010}}.
  See also references therein.

\bibitem{opal}
\hrefCMSnoop {} {{ OPAL} Collaboration, ``Search for chargino and neutralino
  production at $\sqrt{s} = 192-209$ GeV at LEP'',} \textit{ Eur. Phys. J. C}
  \textbf{ 35} (2004) 1,
  \href{http://dx.doi.org/10.1140/epjc/s2004-01758-8}{\doi{10.1140/epjc/s2004-01758-8}}.
  See also references therein.

\bibitem{Aaltonen:2008pv}
\hrefCMSnoop {} {{ CDF} Collaboration, ``{Search for supersymmetry in
  $\textrm{p} \bar{\textrm{p}}$ collisions at $\sqrt{s}$ = 1.96-TeV using the
  trilepton signature for chargino-neutralino production}'',} \textit{ Phys.
  Rev. Lett.} \textbf{ 101} (2008) 251801,
\href{http://dx.doi.org/10.1103/PhysRevLett.101.251801}{\doi{10.1103/PhysRevLett.101.251801}}.
%%CITATION = 0808.2446;%%.

\bibitem{Abazov:2009zi}
\hrefCMSnoop {} {{ D0} Collaboration, ``{Search for associated production of
  charginos and neutralinos in the trilepton final state using 2.3 fb$^{-1}$ of
  data}'',} \textit{ Phys. Lett. B} \textbf{ 680} (2009) 34,
\href{http://dx.doi.org/10.1016/j.physletb.2009.08.011}{\doi{10.1016/j.physletb.2009.08.011}}.
%%CITATION = 0901.0646;%%.

\bibitem{atlas-trilepton}
\hrefCMSnoop {} {{ ATLAS} Collaboration, ``{Search for supersymmetry in events
  with three leptons and missing transverse momentum in $\sqrt{s}$ = 7 TeV pp
  collisions with the ATLAS detector}'',} \textit{ Phys. Rev. Lett.} \textbf{
  108} (2012) 261804,
  \href{http://dx.doi.org/10.1103/PhysRevLett.108.261804}{\doi{10.1103/PhysRevLett.108.261804}}.

\bibitem{Chatrchyan:2011ff}
\hrefCMSnoop {} {{ CMS} Collaboration, ``Search for physics beyond the standard
  model using multilepton signatures in pp collisions at $\sqrt{s}$ = 7 TeV'',}
  \textit{ Phys. Lett. B} \textbf{ 704} (2011) 411,
  \href{http://dx.doi.org/10.1016/j.physletb.2011.09.047}{\doi{10.1016/j.physletb.2011.09.047}}.

\bibitem{SUS-11-013-paper}
\hrefCMSnoop {} {{ CMS} Collaboration, ``{Search for anomalous production of
  multilepton events in pp collisions at $\sqrt{s}=7$ TeV}'',} \textit{ JHEP}
  \textbf{ 06} (2012) 169,
  \href{http://dx.doi.org/10.1007/JHEP06(2012)169}{\doi{10.1007/JHEP06(2012)169}}.

\bibitem{SUS-11-021-paper}
\hrefCMSnoop {} {{ CMS} Collaboration, ``Search for physics beyond the standard
  model in events with a Z boson, jets, and missing transverse energy in pp
  collisions at $\sqrt{s}$ = 7 TeV'',} (2012).
  \href{http://www.arXiv.org/abs/1204.3774}{\texttt{ arXiv:1204.3774}}.
Accepted by \textit{Phys. Lett. B}.
%%CITATION = ARXIV:1204.3774;%%.

\bibitem{SUS-11-010}
\hrefCMSnoop {} {{ CMS} Collaboration, ``{Search for new physics with same-sign
  isolated dilepton events with jets and missing transverse energy}'',} (2012).
  \href{http://www.arXiv.org/abs/1205.6615}{\texttt{ arXiv:1205.6615}}.
Submitted to \textit{Phys. Rev. Lett.}
%%CITATION = ARXIV:1205.6615;%%.

\bibitem{Knuteson:2006ha}
\hrefCMSnoop {} {B.~Knuteson and S.~Mrenna, ``{BARD: interpreting new frontier
  energy collider physics}'',} (2006).
\href{http://www.arXiv.org/abs/hep-ph/0602101}{\texttt{ arXiv:hep-ph/0602101}}.
%%CITATION = HEP-PH/0602101;%%.

\bibitem{ArkaniHamed:2007fw}
N.~Arkani-Hamed\hrefCMSnoop {} { {et~al.}, ``{MARMOSET}: the path from {LHC}
  data to the new standard model via on-shell effective theories'',} (2007).
  \href{http://www.arXiv.org/abs/hep-ph/0703088}{\texttt{
  arXiv:hep-ph/0703088}}.

\bibitem{Dube:2008kf}
S.~Dube\hrefCMSnoop {} { {et~al.}, ``{Addressing the multi-channel inverse
  problem at high energy colliders: a model-independent approach to the search
  for new physics with trileptons}'',} \textit{ J. Phys. G} \textbf{ 39} (2012)
  085004,
\href{http://dx.doi.org/10.1088/0954-3899/39/8/085004}{\doi{10.1088/0954-3899/39/8/085004}}.
%%CITATION = ARXIV:0808.1605;%%.

\bibitem{Alwall:2008ag}
\hrefCMSnoop {} {J.~Alwall, P.~Schuster, and N.~Toro, ``Simplified models for a
  first characterization of new physics at the {LHC}'',} \textit{ Phys. Rev. D}
  \textbf{ 79} (2009) 075020,
\href{http://dx.doi.org/10.1103/PhysRevD.79.075020}{\doi{10.1103/PhysRevD.79.075020}}.
%%CITATION = 0810.3921;%%.

\bibitem{Alwall:2008va}
J.~Alwall\hrefCMSnoop {} { {et~al.}, ``Model-independent jets plus missing
  energy searches'',} \textit{ Phys. Rev. D} \textbf{ 79} (2009) 015005,
  \href{http://dx.doi.org/10.1103/PhysRevD.79.015005}{\doi{10.1103/PhysRevD.79.015005}}.

\bibitem{Alves:2011wf}
\hrefCMSnoop {} {D.~Alves {et~al.}, ``Simplified models for {LHC} new physics
  searches'',} (2011). \href{http://www.arXiv.org/abs/1105.2838}{\texttt{
  arXiv:1105.2838}}.

\bibitem{Alves:2011sq}
\hrefCMSnoop {} {D.~S.~M. Alves, E.~Izaguirre, and J.~G. Wacker, ``Where the
  sidewalk ends: jets and missing energy search strategies for the 7 {TeV
  LHC}'',} \textit{ JHEP} \textbf{ 10} (2011) 012,
\href{http://dx.doi.org/10.1007/JHEP10(2011)012}{\doi{10.1007/JHEP10(2011)012}}.
%%CITATION = ARXIV:1102.5338;%%.

\bibitem{Papucci:20113g}
\hrefCMSnoop {} {M.~Papucci, J.~Ruderman, and A.~Weiler, ``{Natural SUSY
  endures}'',} (2011).
\href{http://www.arXiv.org/abs/1110.6926}{\texttt{ arXiv:1110.6926}}.
%%CITATION = 1110.6926;%%.

\bibitem{Matchev:1999ft}
\hrefCMSnoop {} {K.~T. Matchev and S.~D. Thomas, ``{Higgs and $Z$ boson
  signatures of supersymmetry}'',} \textit{ Phys. Rev. D} \textbf{ 62} (2000)
  077702,
\href{http://dx.doi.org/10.1103/PhysRevD.62.077702}{\doi{10.1103/PhysRevD.62.077702}}.
%%CITATION = HEP-PH/9908482;%%.

\bibitem{Meade:2009qv}
\hrefCMSnoop {} {P.~Meade, M.~Reece, and D.~Shih, ``{Prompt decays of general
  neutralino NLSPs at the Tevatron}'',} \textit{ JHEP} \textbf{ 05} (2010) 105,
\href{http://dx.doi.org/10.1007/JHEP05(2010)105}{\doi{10.1007/JHEP05(2010)105}}.
%%CITATION = ARXIV:0911.4130;%%.

\bibitem{ref:ewkino}
\hrefCMSnoop {} {J.~T. Ruderman and D.~Shih, ``{General neutralino NLSPs at the
  early LHC}'',} (2011).
\href{http://www.arXiv.org/abs/1103.6083}{\texttt{ arXiv:1103.6083}}.
%%CITATION = ARXIV:1103.6083;%%.

\bibitem{Matchev:1999nb}
\hrefCMSnoop {} {K.~T. Matchev and D.~M. Pierce, ``{Supersymmetry reach of the
  Tevatron via trilepton, like sign dilepton and dilepton plus $\tau$ jet
  signatures}'',} \textit{ Phys. Rev. D} \textbf{ 60} (1999) 075004,
\href{http://dx.doi.org/10.1103/PhysRevD.60.075004}{\doi{10.1103/PhysRevD.60.075004}}.
%%CITATION = HEP-PH/9904282;%%.

\bibitem{Baer:1999bq}
H.~Baer\hrefCMSnoop {} { {et~al.}, ``{Trilepton signal for supersymmetry at the
  Fermilab Tevatron revisited}'',} \textit{ Phys. Rev. D} \textbf{ 61} (2000)
  095007,
\href{http://dx.doi.org/10.1103/PhysRevD.61.095007}{\doi{10.1103/PhysRevD.61.095007}}.
%%CITATION = HEP-PH/9906233;%%.

\bibitem{Matchev:1999yn}
\hrefCMSnoop {} {K.~T. Matchev and D.~M. Pierce, ``{New backgrounds in
  trilepton, dilepton and dilepton plus $\tau$ jet SUSY signals at the
  Tevatron}'',} \textit{ Phys. Lett. B} \textbf{ 467} (1999) 225,
\href{http://dx.doi.org/10.1016/S0370-2693(99)01155-7}{\doi{10.1016/S0370-2693(99)01155-7}}.
%%CITATION = HEP-PH/9907505;%%.

\bibitem{Barger:1998hp}
\hrefCMSnoop {} {V.~D. Barger and C.~Kao, ``{Trilepton signature of minimal
  supergravity at the upgraded Tevatron}'',} \textit{ Phys. Rev. D} \textbf{
  60} (1999) 115015,
\href{http://dx.doi.org/10.1103/PhysRevD.60.115015}{\doi{10.1103/PhysRevD.60.115015}}.
%%CITATION = HEP-PH/9811489;%%.

\bibitem{atlas1208.3144}
\hrefCMSnoop {} {{ ATLAS} Collaboration, ``Search for direct production of
  charginos and neutralinos in events with three leptons and missing transverse
  momentum in $\sqrt{s}$ = 7 TeV $pp$ collisions with the {ATLAS} detector'',}
  (2012). \href{http://www.arXiv.org/abs/1208.3144}{\texttt{ arXiv:1208.3144}}.
  Submitted to \textit{Phys. Lett. B}.

\bibitem{atlas1208.2884}
\hrefCMSnoop {} {{ ATLAS} Collaboration, ``Search for direct slepton and
  gaugino production in final states with two leptons and missing transverse
  momentum with the {ATLAS} detector in $pp$ collisions at $\sqrt{s}$ = 7
  TeV'',} (2012). \href{http://www.arXiv.org/abs/1208.2884}{\texttt{
  arXiv:1208.2884}}. Submitted to \textit{Phys. Lett. B}.

\bibitem{:2008zzk}
\hrefCMSnoop {} {{ CMS} Collaboration, ``The {CMS} experiment at the {CERN}
  {LHC}'',} \textit{ JINST} \textbf{ 3} (2008) S08004,
\href{http://dx.doi.org/10.1088/1748-0221/3/08/S08004}{\doi{10.1088/1748-0221/3/08/S08004}}.
%%CITATION = JINST,3,S08004;%%.

\bibitem{Maltoni:2002qb}
\hrefCMSnoop {} {F.~Maltoni and T.~Stelzer, ``{MadEvent: automatic event
  generation with MadGraph}'',} \textit{ JHEP} \textbf{ 02} (2003) 027,
\href{http://dx.doi.org/10.1088/1126-6708/2003/02/027}{\doi{10.1088/1126-6708/2003/02/027}}.
%%CITATION = HEP-PH/0208156;%%.

\bibitem{Alwall:2011uj}
J.~Alwall\hrefCMSnoop {} { {et~al.}, ``{MadGraph} 5: going beyond'',} \textit{
  JHEP} \textbf{ 06} (2011) 128,
  \href{http://dx.doi.org/10.1007/JHEP06(2011)128}{\doi{10.1007/JHEP06(2011)128}}.

\bibitem{Sjostrand:2007gs}
\hrefCMSnoop {} {T.~Sj{\"o}strand, S.~Mrenna, and P.~Z. Skands, ``{A brief
  introduction to PYTHIA 8.1}'',} \textit{ Comput. Phys. Commun.} \textbf{ 178}
  (2008) 852,
\href{http://dx.doi.org/10.1016/j.cpc.2008.01.036}{\doi{10.1016/j.cpc.2008.01.036}}.
%%CITATION = 0710.3820;%%.

\bibitem{PhysRevD.78.013004}
P.~M. Nadolsky\hrefCMSnoop {} { {et~al.}, ``Implications of CTEQ global
  analysis for collider observables'',} \textit{ Phys. Rev. D} \textbf{ 78}
  (2008) 013004,
  \href{http://dx.doi.org/10.1103/PhysRevD.78.013004}{\doi{10.1103/PhysRevD.78.013004}}.

\bibitem{Campbell:2011bn}
\hrefCMSnoop {} {J.~M. Campbell, R.~Ellis, and C.~Williams, ``{Vector boson
  pair production at the LHC}'',} \textit{ JHEP} \textbf{ 07} (2011) 018,
  \href{http://dx.doi.org/10.1007/JHEP07(2011)018}{\doi{10.1007/JHEP07(2011)018}}.

\bibitem{Beenakker:1999xh}
W.~Beenakker\hrefCMSnoop {} { {et~al.}, ``Production of charginos, neutralinos,
  and sleptons at hadron colliders'',} \textit{ Phys. Rev. Lett.} \textbf{ 83}
  (1999) 3780,
\href{http://dx.doi.org/10.1103/PhysRevLett.83.3780}{\doi{10.1103/PhysRevLett.83.3780}}.
%%CITATION = HEP-PH/9906298;%%.

\bibitem{Beenakker:1999xhErr}
W.~Beenakker\hrefCMSnoop {} { {et~al.}, ``Erratum: production of charginos,
  neutralinos, and sleptons at hadron colliders'',} \textit{ Phys. Rev. Lett.}
  \textbf{ 100} (2008) 029901,
\href{http://dx.doi.org/10.1103/PhysRevLett.100.029901}{\doi{10.1103/PhysRevLett.100.029901}}.
%%CITATION = HEP-PH/9906298;%%.

\bibitem{bib-NLO-NLL}
M.~Kr\"{a}mer\hrefCMSnoop {} { {et~al.}, ``Supersymmetry production cross
  sections in pp collisions at $\sqrt{s}=7$~{TeV}'',} (2012).
\href{http://www.arXiv.org/abs/1206.2892}{\texttt{ arXiv:1206.2892}}.
%%CITATION = arXiv:1206.2892;%%.

\bibitem{Skands:2003cj}
\hrefCMSnoop {} {P.~Z. Skands {et~al.}, ``{SUSY Les Houches accord: interfacing
  SUSY spectrum calculators, decay packages, and event generators}'',} \textit{
  JHEP} \textbf{ 07} (2004) 036,
  \href{http://dx.doi.org/10.1088/1126-6708/2004/07/036}{\doi{10.1088/1126-6708/2004/07/036}}.

\bibitem{Baer:1993ae}
H.~Baer\hrefCMSnoop {} { {et~al.}, ``{Simulating supersymmetry with ISAJET 7.0
  / ISASUSY 1.0}'',} (1993).
  \href{http://www.arXiv.org/abs/hep-ph/9305342}{\texttt{
  arXiv:hep-ph/9305342}}.

\bibitem{Beenakker:1996ed}
\hrefCMSnoop {} {W.~Beenakker, R.~Hoepker, and M.~Spira, ``{PROSPINO: a program
  for the production of supersymmetric particles in next-to-leading order
  QCD}'',} (1996).
\href{http://www.arXiv.org/abs/hep-ph/9611232}{\texttt{ arXiv:hep-ph/9611232}}.
%%CITATION = HEP-PH/9611232 ;%%.

\bibitem{Abdullin:1328345}
\hrefCMSnoop {} {{ CMS} Collaboration, ``The fast simulation of the {CMS}
  Detector at the {LHC}'',} in \textit{ International Conference on Computing
  in High Energy and Nuclear Physics (CHEP 2010)}.
\newblock \textit{J. Phys.: Conference Series} 331 (2011) 032049.
  \href{http://dx.doi.org/10.1088/1742-6596/331/3/032049}{\doi{10.1088/1742-6596/331/3/032049}}.

\bibitem{Geant}
\hrefCMSnoop {} {S.~Agostinelli {et~al.}, ``{GEANT4} -- a simulation
  toolkit'',} \textit{ Nucl. Instr. Meth. A} \textbf{ 506} (2003) 250,
\href{http://dx.doi.org/10.1016/S0168-9002(03)01368-8}{\doi{10.1016/S0168-9002(03)01368-8}}.
%%CITATION = NUIMA,A506,250;%%.

\bibitem{PFT-10-004}
\href {http://cdsweb.cern.ch/record/1279358} {{ CMS} Collaboration, ``Study of
  tau reconstruction algorithms using {\Pp\Pp} collisions data collected at
  $\sqrt{s} = 7\,${TeV}'',} CMS Physics Analysis Summary CMS-PAS-PFT-10-004,
  (2010).

\bibitem{PFT-08-001}
\href {http://cdsweb.cern.ch/record/1198228} {{ CMS} Collaboration, ``CMS
  strategies for tau reconstruction and identification using particle-flow
  techniques'',} CMS Physics Analysis Summary CMS-PAS-PFT-08-001, (2009).

\bibitem{EGM-10-004}
\href {http://cdsweb.cern.ch/record/1299116} {{ CMS} Collaboration, ``Electron
  reconstruction and identification at $\sqrt{s} = 7$ {TeV}'',} CMS Physics
  Analysis Summary CMS-PAS-EGM-10-004, (2010).

\bibitem{MUO-10-004}
\hrefCMSnoop {} {{ CMS} Collaboration, ``{Performance of {CMS} muon
  reconstruction in pp collision events at $\sqrt{s}$ = 7 TeV}'',} (2012).
  \href{http://www.arXiv.org/abs/1206.4071}{\texttt{ arXiv:1206.4071}}.
Submitted to \textit{JINST}.
%%CITATION = ARXIV:1206.4071;%%.

\bibitem{PFT-10-XXX}
\hrefCMSnoop {} {{ CMS} Collaboration, ``{Performance of $\tau$ lepton
  reconstruction and identification in CMS}'',} \textit{ JINST} \textbf{ 7}
  (2012) P01001,
\href{http://dx.doi.org/10.1088/1748-0221/7/01/P01001}{\doi{10.1088/1748-0221/7/01/P01001}}.
%%CITATION = JINST,7,P01001;%%.

\bibitem{Cacciari:2008gp}
\hrefCMSnoop {} {M.~Cacciari, G.~P. Salam, and G.~Soyez, ``{The anti-$k_t$ jet
  clustering algorithm}'',} \textit{ JHEP} \textbf{ 04} (2008) 063,
  \href{http://dx.doi.org/10.1088/1126-6708/2008/04/063}{\doi{10.1088/1126-6708/2008/04/063}}.

\bibitem{btag-11-001}
\hrefCMSnoop {} {{ CMS} Collaboration, ``{Performance of the b-jet
  identification in CMS}'',} CMS Physics Analysis Summary CMS-PAS-BTV-11-001,
  (2011).

\bibitem{WZpaper}
\hrefCMSnoop {} {{ CMS} Collaboration, ``Measurement of the inclusive W and Z
  production cross sections in pp collisions at $\sqrt {s} = 7 $ TeV with the
  CMS experiment'',} \textit{ JHEP} \textbf{ 10} (2011) 132,
  \href{http://dx.doi.org/10.1007/JHEP10(2011)132}{\doi{10.1007/JHEP10(2011)132}}.

\bibitem{Nachtman:1999ua}
\hrefCMSnoop {} {J.~Nachtman, D.~Saltzberg, and M.~Worcester, ``{Study of a
  like sign dilepton search for chargino neutralino production at CDF}'',}
  (1999). \href{http://www.arXiv.org/abs/hep-ex/9902010}{\texttt{
  arXiv:hep-ex/9902010}}.
FERMILAB-CONF-99-023-E.
%%CITATION = HEP-EX/9902010;%%.

\bibitem{Lykken:1999kp}
\hrefCMSnoop {} {J.~D. Lykken and K.~T. Matchev, ``{Supersymmetry signatures
  with $\tau$ jets at the Tevatron}'',} \textit{ Phys. Rev. D} \textbf{ 61}
  (2000) 015001,
\href{http://dx.doi.org/10.1103/PhysRevD.61.015001}{\doi{10.1103/PhysRevD.61.015001}}.
%%CITATION = HEP-PH/9903238;%%.

\bibitem{Junk:1999kv}
\hrefCMSnoop {} {T.~Junk, ``{Confidence level computation for combining
  searches with small statistics}'',} \textit{ Nucl. Instrum. Meth. A} \textbf{
  434} (1999) 435,
\href{http://dx.doi.org/10.1016/S0168-9002(99)00498-2}{\doi{10.1016/S0168-9002(99)00498-2}}.
%%CITATION = HEP-EX/9902006;%%.

\bibitem{Read:2002hq}
\hrefCMSnoop {} {A.~L. Read, ``{Presentation of search results: The $CL_s$
  technique}'',} \textit{ J. Phys. G} \textbf{ 28} (2002) 2693,
\href{http://dx.doi.org/10.1088/0954-3899/28/10/313}{\doi{10.1088/0954-3899/28/10/313}}.
%%CITATION = JPHGB,G28,2693;%%.

\bibitem{ATLAS:1379837}
\href {http://cdsweb.cern.ch/record/1379837} {{ATLAS and CMS Collaborations},
  ``Procedure for the {LHC} Higgs boson search combination in summer 2011'',}
  Technical Report ATL-PHYS-PUB-2011-11, CMS-NOTE-2011-005, Geneva, (2011).

\bibitem{PDG}
\hrefCMSnoop {} {{Particle Data Group}, J.~Beringer, {et~al.}, ``{Review of
  particle physics}'',} \textit{ Phys. Rev. D} \textbf{ 86} (2012) 010001,
  \href{http://dx.doi.org/10.1103/PhysRevD.86.010001}{\doi{10.1103/PhysRevD.86.010001}}.

\end{thebibliography}\endgroup

\clearpage
\appendix

\section{Signal efficiency model for the three-lepton analysis
with \texorpdfstring{$\mdil$}{dilepton mass}
and \texorpdfstring{$\MT$}{transverse mass}
}
\label{efficiency}

In order to facilitate the interpretation of the three-lepton results
with $\mdil$ and $\MT$ presented in Section~\ref{trilepton-targeted}
within the context of other signal models that are not considered
here, we provide a prescription for emulating the event selection
efficiency.  This prescription includes lepton reconstruction and
identification efficiencies, $\MET$ and $\MT$ selection efficiencies,
as well as the \bjet identification probability.  The latter can be
used to parameterize the $\cPqb$-veto acceptance in case the model of
interest contains such jets.

We perform a fit to efficiency curves for each selection using the
parametric function
\begin{equation}\label{eq:EffFunction}
\epsilon (x) = p_{6} + p_{4} \left[ \rm{erf}\left( \frac{ x - p_{0} }{ p_{1} } \right) + 1 \right]  +
                       p_{5} \left[ \rm{erf}\left( \frac{ x - p_{2} }{ p_{3} } \right) + 1 \right]  ,
\end{equation}
where $x$ represents the observable for which the efficiency is
parametrized, and $\rm{erf}$ indicates the error function.  This
includes the efficiency for electrons and muons to be reconstructed
and to satisfy the identification requirements as a function of the
lepton $\PT$; the probability for an event to satisfy the requirements
$\MET > 50\GeV$ and $\MT > 100\GeV$ as a function of true $\MET$
and true $\MT$; and the probability for a jet to be identified as a
\bjetnohyphen separately for the cases where the jet originates from a
$\cPqb$-, $\cPqc$-, or light-flavor quark or gluon as a function of
jet $\PT$.  (The true $\MET$ observable is calculated with the stable
generator-level invisible particles, while the true $\MT$ is
calculated using the true $\MET$ and the third lepton, \ie, the one
not used in the $\mdil$ calculation.)

The parameters of the fitted functions are given in
Table~\ref{tab:par}.  Using these parameters and the values of $x$, a
combined probability for a given event to pass the full event
selection can be obtained. We have tested the efficiency model in a
signal sample and observed consistent event yields compared to the
full detector simulation within about 25\%.

\begin{table}[h]
\begin{center}
\topcaption{ The parameters of the efficiency function $\epsilon (x)$,
where $x$ represents $\PT(\mu)$, $\PT(\Pe)$, $\MET$, $\MT$, or
$\PT(\text{parton})$ for different quark flavors
($\cPqu\cPqd\cPqs\cPqc\cPqb$) and for gluons ($\cPg$).
}
\label{tab:par}
\begin{tabular}{lccccccc} \hline
 $x$ &  $p_{0}$  & $p_{1}$  &  $p_{2}$ &  $p_{3}$   & $p_{4}$ & $p_{5} $ & $p_{6} $	 \\ \hline  \hline
$\PT(\mu)$  &-4.65      & 27.38    &-14.64    &-9.31      & 0.47     &-849.3    & 0.\\
$\PT(\Pe)$,   &12.32      &10.11     &20.12     &32.17      & 0.32     & 0.11     & 0.\\
$\MET$   &48.37     &43.54     &49.90     &14.95      & 0.06     & 0.44     & 0.\\
$\MT$     &98.23     &87.99     &97.61     &29.78      & 0.36     & 0.14     & 0.008\\
$\PT(\bquark)$        &30.60 & 31.80 & 0.34& 0.& 0.& 0.& 0.\\
$\PT(\cquark)$        &32.02 & 45.34 & 0.11& 0.& 0.& 0.& 0.\\
$\PT(\udsgparton)$    &68.84 & 55.21 & 0.02& 0.& 0.& 0.& 0.\\ \hline
\end{tabular}
\end{center}
\end{table}

\cleardoublepage \appendix\section{The CMS Collaboration \label{app:collab}}\begin{sloppypar}\hyphenpenalty=5000\widowpenalty=500\clubpenalty=5000\textbf{Yerevan Physics Institute,  Yerevan,  Armenia}\\*[0pt]
S.~Chatrchyan, V.~Khachatryan, A.M.~Sirunyan, A.~Tumasyan
\vskip\cmsinstskip
\textbf{Institut f\"{u}r Hochenergiephysik der OeAW,  Wien,  Austria}\\*[0pt]
W.~Adam, E.~Aguilo, T.~Bergauer, M.~Dragicevic, J.~Er\"{o}, C.~Fabjan\cmsAuthorMark{1}, M.~Friedl, R.~Fr\"{u}hwirth\cmsAuthorMark{1}, V.M.~Ghete, J.~Hammer, N.~H\"{o}rmann, J.~Hrubec, M.~Jeitler\cmsAuthorMark{1}, W.~Kiesenhofer, V.~Kn\"{u}nz, M.~Krammer\cmsAuthorMark{1}, I.~Kr\"{a}tschmer, D.~Liko, I.~Mikulec, M.~Pernicka$^{\textrm{\dag}}$, B.~Rahbaran, C.~Rohringer, H.~Rohringer, R.~Sch\"{o}fbeck, J.~Strauss, A.~Taurok, W.~Waltenberger, G.~Walzel, E.~Widl, C.-E.~Wulz\cmsAuthorMark{1}
% ...
\vskip\cmsinstskip
\dag:~Deceased\\
1:~~Also at Vienna University of Technology, Vienna, Austria\\
% ...

\end{sloppypar}
\end{document}